\documentclass[a4paper,11pt]{article}
\usepackage[utf8]{inputenc}
\usepackage{graphicx}
\usepackage{hyperref}
\usepackage{latexsym}
\usepackage{color}
\usepackage{amsmath}
\usepackage{amssymb}
\usepackage[total={15cm,25cm}]{geometry}
\usepackage{slashed}

\def\msbar{{\overline{\rm MS}}}

\renewcommand\arraystretch{1.8}

\title{Charm physics with overlap fermions on 2+1-flavor domain wall fermion configurations}
\author{Dong-Hao Li$^{1,2,3}$\thanks{lidonghao@ihep.ac.cn}, Ying Chen$^{2,3}$\thanks{cheny@ihep.ac.cn}, Ming Gong$^{2,3}$\thanks{gongming@ihep.ac.cn}, Keh-Fei Liu$^{4,5}$\thanks{liu@g.uky.edu},\\
Zhaofeng Liu$^{2,6}$\thanks{liuzf@ihep.ac.cn, corresponding author}, Ting-Xiao Wang$^{2,3}$\thanks{wangtx@ihep.ac.cn, corresponding author}\\
($\chi$QCD Collaboration)}
\date{}

\begin{document}

\maketitle
\begin{center}
$^1$MOE Frontiers Science Center for Rare Isotopes, \\and School of Nuclear Science and Technology, Lanzhou University, Lanzhou 730000, China \\
$^2$Institute of High Energy Physics, Chinese Academy of Sciences, Beijing 100049, China\\
$^3$School of Physical Sciences, University of Chinese Academy of Sciences, Beijing 101408, China\\
$^4$Department of Physics and Astronomy, University of Kentucky, Lexington, Kentucky 40506, USA\\
$^5$Nuclear Science Division, Lawrence 
Berkeley National Laboratory, Berkeley, California 94720, USA\\
$^6$Center for High Energy Physics, Peking University, Beijing 100871, China
\end{center}

\begin{abstract}
Decay constants of pseudoscalar mesons $D$, $D_s$, $\eta_c$ and vector mesons $D^*$, $D_s^*$, $J/\psi$ are determined from $N_f=2+1$ lattice QCD at a lattice spacing $a\sim0.08$ fm. For vector mesons, the decay constants defined by tensor currents are given in the $\msbar$ scheme at $2$ GeV.
The calculation is performed on domain wall fermion configurations generated by the RBC-UKQCD Collaborations and the overlap fermion action is used for the valence quarks. Comparing the current results with our previous ones at a coarser lattice spacing $a\sim0.11$ fm gives us a better understanding of the discretization error.
We obtain $f_{D_s^*}^T(\msbar,\mbox{ 2 GeV})/f_{D_s^*}=0.909(18)$ with a better precision than our previous result. Combining our $f_{D_s^*}=277(11)$ MeV with the total width of $D_s^*$ determined in a recent work gives a branching fraction $4.26(52)\times10^{-5}$ for $D_s^*$ leptonic decay.

\end{abstract}

\section{Introduction}
\label{sec:intro}
Charm physics provides a rich phenomenology and offers us a valuable platform for precisely testing the Standard Model and understanding low-energy Quantum Chromodynamics (QCD). 
The decay constants of charmed pseudoscalar mesons combined with experimental data of the relevant leptonic decays can be used to extract the Cabibbo-Kobayashi-Maskawa (CKM) matrix elements $V_{cd}$ and $V_{cs}$ (see \cite{ParticleDataGroup:2022pth}), or serve as probes for new particles of new physics such as charged Higgs bosons (see \cite{Dobrescu:2008er}). 
However, perturbative theory and heavy quark expansion are rather precarious in charm sector, and our most reliable theoretical tool is lattice QCD.
An overview of the decay constants of charmed pseudoscalar mesons can be found in Ref.~\cite{FlavourLatticeAveragingGroupFLAG:2021npn} for lattice QCD calculations before 2022 (see the individual results in Refs.~\cite{FermilabLattice:2011njy,Na:2012iu,Carrasco:2014poa,Boyle:2017jwu,Bazavov:2017lyh,Boyle:2018knm}). 
Two recent $2+1$-flavor calculations of these decay constants appeared in Refs.~\cite{Bussone:2023kag,Kuberski:2024pms}.

For the pseudoscalar meson $D$ or $D_s$, its pure leptonic decay $D_{(s)}\to \ell\nu_\ell$, of which the decay width is proportional to $f^2_P|V_{cq}|^2$ ($q = d\mbox{ or }s$ with $f_P$ the decay constant of $D\mbox{ or }D_s$, respectively), provides a clean channel to determine the CKM matrix elements. 
Moreover, the semileptonic processes $D_{(s)}\to h\ell\nu_\ell$, with $h$ representing a pion or kaon, are induced by the vector current $\bar{q}\gamma^\mu c$, while the pure leptonic processes are induced by the axial-vector current $\bar{q}\gamma^\mu\gamma_5 c$.
Comparing the CKM matrix elements extracted from these two kinds of processes allows us to test the $V-A$ structure of Weak interaction. 
The vector mesons $D_{(s)}^*$ almost 100\% decay respectively to the corresponding charmed pseudoscalar  mesons, with relatively small leptonic decay branching ratios. 
The first measurement of $f_{D_s^*}$ from the leptonic decay $D_s^*\to e^+ \nu_e$ is recently reported by BESIII \cite{BESIII:2023zjq}. 
In the phenomenological analysis of the nonleptonic weak decays of charmed vector mesons, the decay constants denoted by $f_V$ are necessary inputs \cite{Cheng:2022mvd,Yang:2022ece}. 
In each of the $b\to c$ induced semileptonic or nonleptonic decays of bottom mesons, the decay constant $f_V$ of the charmed meson in the final state, as well as the decay constant $f_V^T$ defined by the tensor current, appears as nonperturbative inputs \cite{Cui:2023jiw,Zhou:2015jba}. 
Furthermore, the ratio of the decay constants of vector mesons and pseudoscalar mesons $f_V/f_P$ approaches to 1 in the heavy quark limit, which can be used to study the breaking of heavy quark symmetry. Currently, unquenched lattice QCD calculations of decay constants of charmed vector mesons are
relatively sparse \cite{Donald:2013sra,Becirevic:2012ti,Blossier:2018jol,Lubicz:2017asp,Gambino:2019vuo}.
Results from sum rules can be found in, for example, Refs.~\cite{Gelhausen:2013wia,Wang:2015mxa}.

For the charmonium system, the leptonic decay constant of $J/\psi$ can be directly determined experimentally \cite{BESIII:2022wsp}. 
In this work, we calculated the ratios of decay constants $f_{J/\psi}/f_{\eta_c}$ and $f^T_{J/\psi}/f_{J/\psi}$, where the former can be directly used in the amplitude analysis of $B$ meson decays into charmonium states, and the latter will appear in $b\to c$ decays induced by specific new physics operators. 
By combining the decay width $\Gamma(J/\psi\to e^+e^-)$ measured experimentally with the ratio $f_{J/\psi}/f_{\eta_c}$, we can determine the decay constant $f_{\eta_c}$, which is used for the calculation of decay amplitudes $\Gamma(\eta_c\to\gamma^*\gamma)$ and form factors appearing in $\gamma^*\gamma^*\to\eta_c$ processes \cite{Babiarz:2019sfa,Ryu:2018egt,Geng:2013yfa}. 
Theoretically, the decay constants $f_{J/\psi}$ and $f_{\eta_c}$ were calculated in quenched lattice QCD in \cite{Dudek:2006ej}. 
Two-flavor lattice QCD calculations of $f_{J/\psi}$ and $f_{\eta_c}$ are given in \cite{Becirevic:2013bsa} and \cite{Bailas:2018car}. The HPQCD collaboration obtains $f_{J/\psi}$ and/or $f_{\eta_c}$ in $2+1$-flavor \cite{Davies:2010ip,Donald:2012ga}, and $2+1+1$-flavor simulations \cite{Hatton:2020qhk}. The latter work also considers the quenched Quantum Electrodynamics (QED) effects of charm quarks.

Our previous $2+1$-flavor calculation \cite{Chen:2020qma} of the decay constants of charmed mesons were done at an inverse lattice spacing $a^{-1} = 1.730(4)\,$GeV. The discretization error was estimated to be about 2\%. We now present the results on a finer lattice with $a^{-1} = 2.383(9)\,$GeV to better understand the lattice cutoff effects. 

In the following we present the details of our calculation framework in Sec. \ref{sec:details}, which includes the lattice setup and the computation of two-point correlation functions. Sec. \ref{sec:Mass} and Sec. \ref{sec:fx} show the analyses of the meson masses and decay constants, respectively. The discussions and summary can be found in Sec. \ref{sec:summary}.

\section{Simulation details}
\label{sec:details}
The $2+1$-flavor gauge configurations used in this work were generated by the RBC-UKQCD Collaborations~\cite{RBC:2010qam}.
The dynamical quarks are domain-wall fermions with degenerate light (up and down) quark masses $am_l^{\rm sea}=0.004$, $0.006$, $0.008$ and strange quark mass $am_s^{\rm sea}=0.03$ in lattice units.
The inverse lattice spacing $a^{-1}$ determined in Ref.~\cite{RBC:2014ntl}, as well as the other parameters of the configurations,
is given in Table~\ref{tab:confs}. The spatial extension of the lattice is about $La\sim2.7$ fm. The light sea quark masses $am_l^{\rm sea}$ mentioned above correspond to pion masses $m_\pi^{\rm sea}$ around $302$, 360 and 412 MeV, respectively~\cite{RBC:2010qam}. More information about the configurations can be found in Ref.~\cite{RBC:2010qam}.
\begin{table}[t]
    \caption{ Configurations used in this work. The residual mass of the dynamical fermion $am_{\rm res}$
is in the two-flavor chiral limit from Ref.~\cite{RBC:2010qam}.
$N_{\rm conf}$ is the number of configurations, and $N_{\rm src}$ the number of point sources on each configuration}
    \label{tab:confs}
    \centering
\begin{tabular}{cccccc}
\hline\hline
$a^{-1}$ (GeV) & Label & $am_l^{\rm sea}/am_s^{\rm sea}$ & Volume & $N_{\rm conf}\times N_{\rm src}$ & $am_{\rm res}$ \\
\hline
2.383(9) & \texttt{f004} & 0.004/0.03 & $32^3\times64$ & $628\times1$ & 0.0006664(76) \\
         & \texttt{f006} & 0.006/0.03 & $32^3\times64$ & $42\times16$ & \\
         & \texttt{f008} & 0.008/0.03 & $32^3\times64$ & $49\times16$ & \\
\hline\hline
\end{tabular}
\end{table}

The valence quarks used in this study are overlap fermions. The massless overlap Dirac operator~\cite{Neuberger:1997fp} is defined as
\begin{equation}
D_{\rm ov}(\rho)=1 + \gamma_5 \varepsilon (\gamma_5 D_{\rm w}(\rho)),
\end{equation}
where $\varepsilon$ is the matrix sign function and $D_{\rm w}(\rho)$ is the usual Wilson fermion operator,
except for a negative mass parameter $- \rho = 1/2\kappa -4$ with $\kappa_c < \kappa < 0.25$ and $\kappa_c$ corresponding to a massless Wilson operator. In practice, we use $\kappa = 0.2$ which corresponds to $\rho = 1.5$. The massive overlap Dirac operator is defined as
\begin{eqnarray}
D_m &=& \rho D_{\rm ov} (\rho) + m\, (1 - \frac{D_{\rm ov} (\rho)}{2}) \nonumber\\
       &=& \rho + \frac{m}{2} + (\rho - \frac{m}{2})\, \gamma_5\, \varepsilon (\gamma_5 D_{\rm w}(\rho)),
\end{eqnarray}
To accommodate the SU(3) chiral symmetry, it is usually convenient to use the chirally regulated field
$\hat{\psi} = (1 - \frac{1}{2} D_{\rm ov}) \psi$ in lieu of $\psi$ in the interpolation field and the currents.
This is equivalent to leaving the currents unmodified and instead adopting the effective propagator
\begin{equation}
G \equiv D_{\rm eff}^{-1} \equiv (1 - \frac{D_{\rm ov}}{2}) D^{-1}_m = \frac{1}{D_c + m},
\end{equation}
where $D_c = \frac{\rho D_{\rm ov}}{1 - D_{\rm ov}/2}$ satisfies $\{\gamma_5, D_c\}=0$~\cite{Chiu:1998gp}.

The decay constant $f_P$ of a pseudoscalar meson $P$ is defined through the following matrix element of the axial-vector current $A_\mu(x)=\bar{q}_1(x) \gamma_\mu \gamma_5 q_2(x)$, 
\begin{equation}
    \left\langle 0\left|A_\mu(x)\right| P(p)\right\rangle=\mathrm{i}p_\mu f_P e^{-\mathrm{i}p\cdot x},\label{lqcd decay1}
\end{equation}
where $p_\mu$ is the four-momentum of the meson $P$, with $q_{1,2}$ referring to the spinor fields of the constituent quarks in it. For charmed mesons $D_{(s)}$, the two quarks are charm and light (strange) quarks, and for charmonium $\eta_c$, both of the constituent quarks are charm quarks. In lattice QCD, the matrix element in Eq.~(\ref{lqcd decay1}) can be extracted by calculating the two-point function involving $A_\mu(x)$,
\begin{equation}
C(t)= \sum\limits_{\vec{x}} \langle 0 | \mathcal{O}(\vec{x},x_0)O^{\dagger}(\vec s,s_0)|0\rangle,
\label{eq:lqcd-2pt}
\end{equation}
where $t\equiv x_0-s_0$ is the time displacement in lattice units between the source point $s$ and sink point $x$. The interpolating operators are $\mathcal{O}=\bar{q}_1\Gamma q_2$, where 
$\Gamma=\gamma_\mu \gamma_5$ for 
$\mathcal{O}=A_\mu$.
For two-point functions of charmonia only the connected quark contractions are considered.

In addition, one needs to determine a normalization constant $Z_A$ for $A_\mu(x)$ due to the fact that, with finite lattice spacings, $A_\mu$ is no longer a conserved current in the chiral limit. The numerical determination of $Z_A$ introduces additional uncertainties. 

For chiral lattice fermions such as domain-wall and overlap fermions one can use the partially conserved axial vector current (PCAC) relation to avoid the computation of $Z_A$. One can get $f_P$ from the matrix element of the pseudoscalar density as
\begin{equation}
      (m_1+m_2) \left\langle 0\left|\bar{q}_1(0)  \gamma_5 q_2(0)\right| P(p)\right\rangle=m_P^2 f_P,\label{lqcd decay2}
\end{equation}
where $m_{1,2}$ are the quark masses and $m_P$ is the mass of the pseudoscalar meson. For overlap fermions the renormalization constants of the quark mass and pseudoscalar density $P=\bar{q}_1\gamma_5 q_2$ cancel each other ($Z_m^{-1}=Z_P$), of which the numerical verification can be found in, for example, Ref.~\cite{DeGrand:2005af}. Then we can use $\Gamma=\gamma_5$ to calculate the two-point function in Eq.~(\ref{eq:lqcd-2pt}) and obtain $f_P$ from Eq.~(\ref{lqcd decay2}).

Vector mesons have two decay constants $f_V$ and $f_V^T$, which are defined by the matrix element of the vector and tensor currents, respectively, between the vacuum and a vector meson $V$ as
\begin{equation}
\begin{aligned}
    \langle 0|\bar q_1(0)\gamma_{\mu}q_2(0)|V(p,\lambda)\rangle &= m_{V}f_V \epsilon_{\mu}(p,\lambda),\\
     \langle 0|\bar q_1(0)\sigma_{\mu\nu} q_2(0)|V(p,\lambda)\rangle &= \mathrm{i} f_V^T \left[\epsilon_{\mu}(p,\lambda) p_\nu - \epsilon_{\nu}(p,\lambda) p_\mu\right],
\label{lqcd:fv}
\end{aligned}
\end{equation}
where $\epsilon_\mu(p,\lambda)$ is the polarization vector of meson $V$ with helicity $\lambda$, and $\sigma_{\mu\nu}=\frac{\mathrm{i}}{2}[\gamma_\mu,\gamma_\nu]$. 
$\Gamma=\{\gamma_i,\sigma_{0i}\}$ are used in Eq.~(\ref{eq:lqcd-2pt}) to extract the decay constants $f_V$ and $f_V^T$, respectively.

The local vector and tensor currents on the lattice should be renormalized by the constants $Z_V$ and $Z_T$, respectively. For overlap fermions we have $Z_V = Z_A$, which was verified numerically in
Refs.~\cite{Liu:2013yxz,Bi:2017ybi,He:2022lse} on various gauge ensembles. The renormalization constants $Z_A$ and $Z_T$ for this work have been calculated in Ref.~\cite{Bi:2023pnf} as
\begin{equation}
    Z_A=1.0789(10), \quad Z_T^{\overline{\mathrm{MS}}}(2 \mathrm{  ~GeV}) / Z_A=1.0721(97), \quad Z_T^{\overline{\mathrm{MS}}}(2 \mathrm{ ~GeV})=1.157(11),
\label{eq:Zfactors}
\end{equation}
where we give values of $Z_T$ in the commonly used $\msbar$ scheme and at a scale $\mu=2$ GeV.

To obtain the two-point functions for various mesons, we calculate quark propagators with a range of masses from the light to charm quark on three ensembles by using Z3-random point sources. The spatial locations of the point sources on ensembles \texttt{f006} and \texttt{f008}
are randomly chosen to reduce correlation. On each of these two ensembles 16 equally-distributed time slices per configuration are used to set the point sources.
On ensemble \texttt{f004} one source at the origin of the lattice on each configuration is used.
In total more than six hundred measurements are done on every ensemble (see Table~\ref{tab:confs}).
    The valence quark masses $am_{l,s,c}$ in lattice units are given in Table~\ref{tab:val}.
\begin{table}[t]
    \caption{ Valence quark masses used in this work.
The physical mass point of valence charm quark is estimated to be around 0.492 in lattice units (see below).}
    \centering
\begin{tabular}{ll}
\hline\hline
   $am_l$ & 0.00460, 0.00585, 0.00677, 0.00765, 0.00885\\
                      & 0.01120, 0.01290, 0.01520, 0.01800, 0.02400 \\
   $m_\pi$ & $\sim$ 220 $-$ 500 MeV  \\
   $am_s$ &   0.037, 0.040, 0.043, 0.046, 0.049, 0.052 \\
   $am_c$  & 0.450, 0.492, 0.500, 0.550   \\
\hline\hline
\end{tabular}
\label{tab:val}
\end{table}


The physical mass points of light, strange and charm quarks are fixed by using the meson masses $m_\pi^{\rm expt}$, $m_K^{\rm expt}$ and $m_{J/\psi}^{\rm expt}$ measured experimentally. However, the corrections from the difference $m_d-m_u$ and the electromagnetic effects should be removed from the experimental values. The corresponding values are defined to be $m_\pi^{\rm phys}\equiv m_{\pi^0}^{\rm expt}$ and ($m_K^{\rm phys})^2\equiv \frac{1}{2}[(m_{K^+}^{\rm latt})^2+(m_{K^0}^{\rm latt})^2]$ in isospin-symmetric 
QCD with $m_{K^+}^{\rm latt}=491.405$ MeV and $m_{K^0}^{\rm latt}=497.567$ MeV  \cite{FermilabLatticeHPQCD:2023jof,Bazavov:2017lyh,Dowdall:2013rya}. We then use $m_\pi^{\rm phys}=134.98$ MeV, $m_K^{\text{phys}}=494.49$ MeV, and $m_{J/\psi}^{\rm phys}=3.0969$ GeV to determine the physical mass points.
Note that at the physical pion mass, $D^*$ decays into the $P$-wave $D\pi$ state through the strong interaction. In Table~\ref{lqcd:open}, we have listed some masses of $D^{(*)}$, with different valence pion masses on our lattice. It is seen that, at each valence pion mass, the mass of the vector meson $D^*$ is below the $D\pi$ threshold. Therefore, in this study $D^*$ is stable on our lattice. Similarly, $D_s^*$ is also a stable particle.
\begin{table}[!htbp]
    \caption{Masses of $D^{(*)}$ in units of GeV, with the corresponding valence pion masses listed, at different valence quark masses.}
    \centering
    \footnotesize
\begin{tabular}{ccccccc}
\hline\hline
   $am_l$ & 0.00460&0.00585& 0.00677& 0.00765&0.00885&$am_c$\\
\hline
   $M_{D^*}$ & 1.937&1.940&1.941&1.943&1.945& 0.450 \\
   $M_{D}$ &   1.767&1.769&1.770&1.772&1.774& 0.450 \\
    $m_{\pi}$  & 0.221&0.250&0.269&0.285&0.307&  \\
\hline\hline
\end{tabular}
\label{lqcd:open}
\end{table}

\section{Data Analyses and results}
\begin{figure}[t]
    \centering
   \includegraphics[width=0.46 \columnwidth]{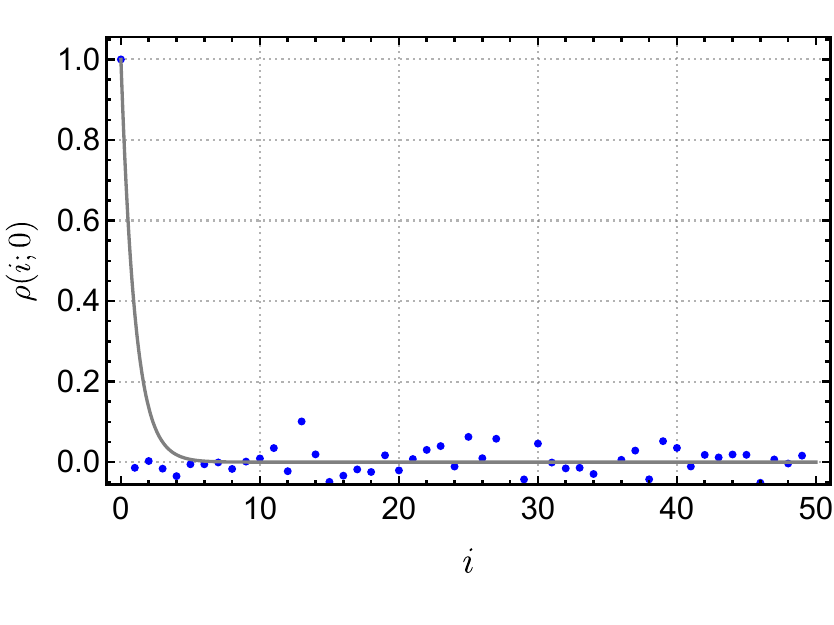}
\includegraphics[width=0.46 \columnwidth]{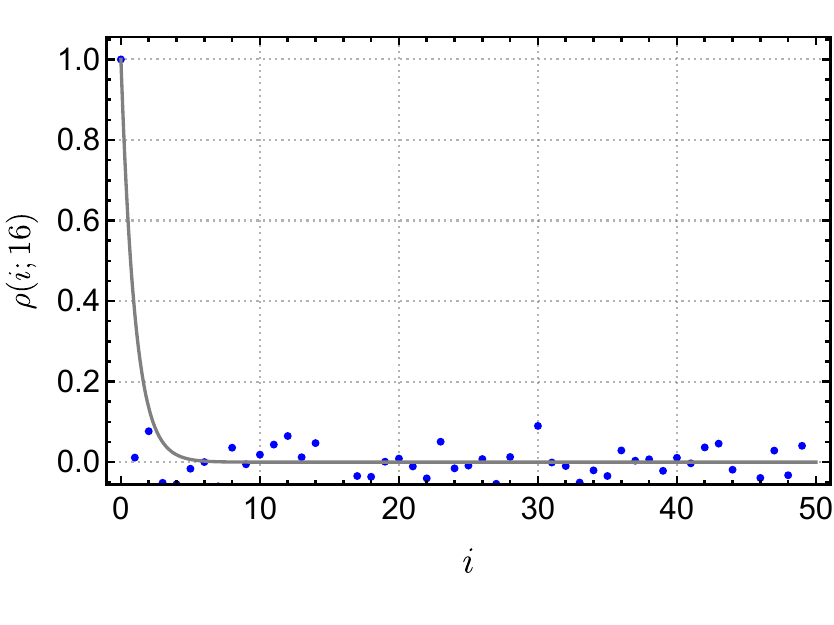}
    \caption{The behavior of autocorrelation functions with measurement interval $i$. The gray lines are schematic exponential functions $e^{-i}$.}
 \label{lqcd fig autocorr}
\end{figure}

As mentioned earlier, 16 sources separated by four time slices are used on each configuration of ensembles \texttt{f006} and \texttt{f008}. We check the autocorrelations among the measurements by computing the normalized autocorrelation function of two-point functions $C(t)$ at several chosen $t$. For example, we compute the autocorrelation function $\rho(i;t)$ of the two-point function $C(t)$ calculated on the \texttt{f008} ensemble with $\mathcal{O}=\bar{s}\gamma_5 c$ and $am_s/am_c=0.037/0.450$
\begin{equation}
\rho(i;t)=\frac{\sum_j[C_j(t)-C(t)][C_{j+i}(t)-C(t)]}{\sum_j[C_j(t)-C(t)]^2},
\end{equation}
where $i,j$ label the measurements.
We choose $t=0,16$, and the variation of $\rho(i;t)$ with $i$ is shown in Fig.~\ref{lqcd fig autocorr}. It is observed that $\rho(i;t)$ decays very rapidly with the measurement separation $i$, indicating little correlation among the measurements of the two-point function. When estimating the statistical errors in our data analyses using the Jackknife resampling method, it is safe enough that we remove 16 measurements from the same configuration on ensemble \texttt{f006} and \texttt{f008} for each resampling. 

We now turn to fitting the two-point function in Eq.~(\ref{eq:lqcd-2pt}), of which the numerical result can be obtained by using the quark propagators. After the intermediate state insertion, the spectral expression of the two-point function under the temporal boundary condition reads 
\begin{equation}
C(t)=\sum_{n=0}^{N_\mathcal{O}-1} A_n \left( e^{-m_n t}+ e^{-m_n(T-t)}\right)\overset{0\ll t\ll T}{\longrightarrow}A_0 \left( e^{-m_0 t}+ e^{-m_0(T-t)}\right),
\label{eq:ct-func0}
\end{equation}
where $N_\mathcal{O}$ denotes the number of all hadron states with the same quantum numbers as the operator $\mathcal{O}$. At large $t$, the contribution from the ground state dominates the two-point function. Thus by fitting $C(t)$ through the function form on the right hand side of Eq.~(\ref{eq:ct-func0}), we can extract the mass $m_0$ of the ground state and the decay constant that is encoded in $A_0$ by the relation $A_0\equiv \left|\langle0|\mathcal{O}|H_0\rangle\right|^2/(2m_0)$. In doing so, we perform a correlated minimal-$\chi^2$ fit and the statistical errors are estimated through the Jackknife analysis.
Practically, we fold the data along $T/2=32$ before the fitting and have taken the following steps to ensure the stability of the fitting result.
\begin{enumerate}
\item Vary $t_{\text{min}}$ for the fitting range $[t_{\text{min}}, T/2]$ to select a stable fitting result.
\item Ensure that $\chi^2/\text{d.o.f}\lesssim 1.0$.
\item 
Use a constrained multi-state fit~\cite{Lepage:2001ym} to obtain the ground state mass $m_0$ again, ensuring that it does not change with the increase of states in the fitting function. 
\item
Verify that the plateau of the effective mass $m_{\text{eff}}$ obtained from $C(t)$ is consistent with the fitted $m_0$ from both the one-state fit and multi-state fit.
\end{enumerate}

\subsection{Meson masses}
\label{sec:Mass}
\begin{figure}[!htbp]
    \centering
   \includegraphics[width=0.32 \columnwidth]{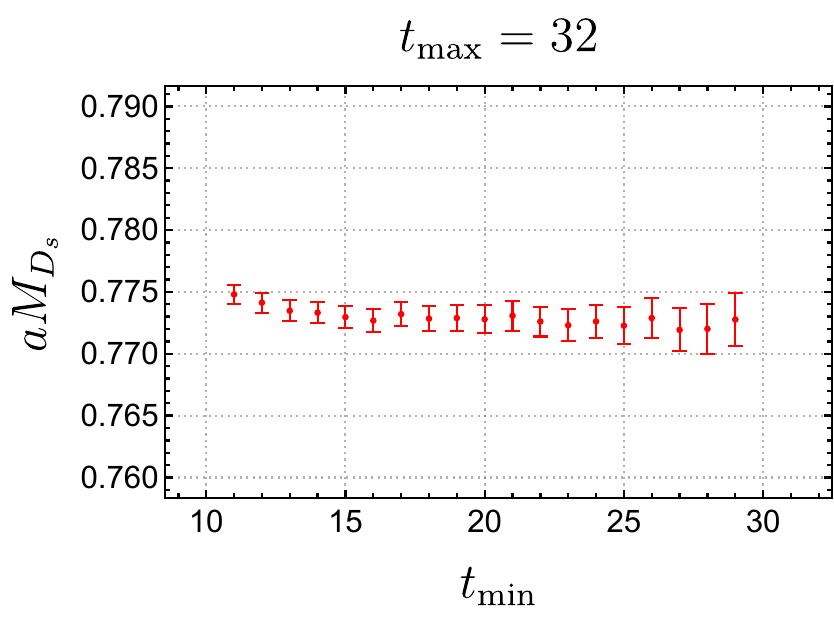}
   \includegraphics[width=0.32 \columnwidth]{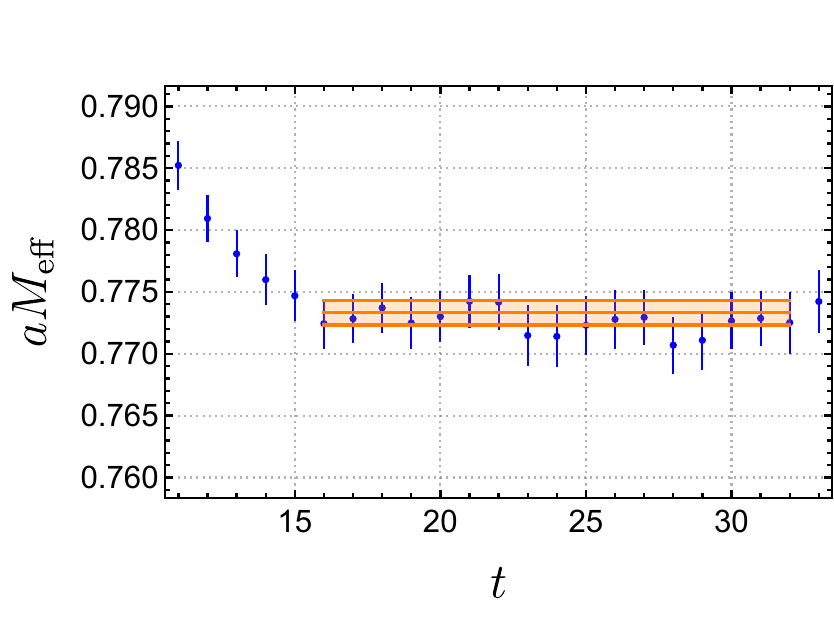}
   \includegraphics[width=0.32 \columnwidth]{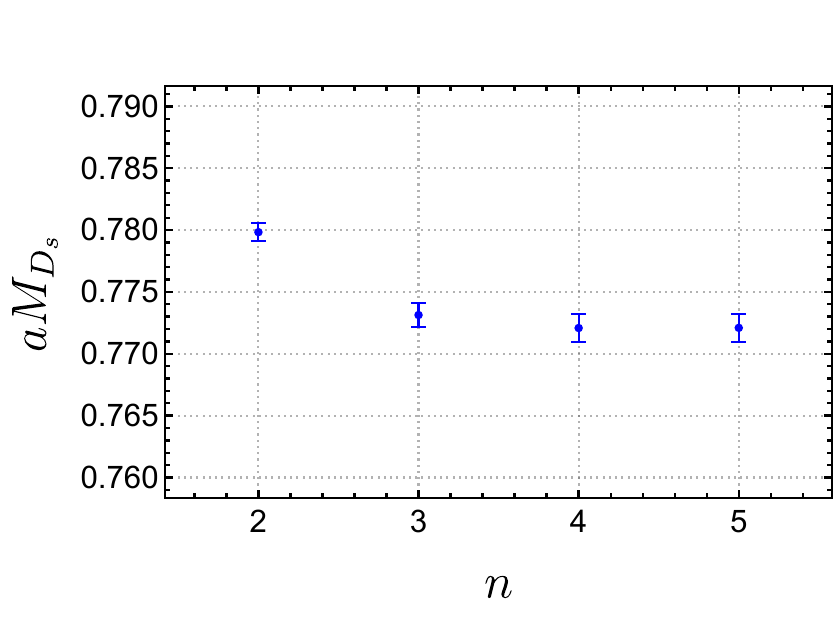}
    \caption{Left: fitted $aM_{D_s}$ as a function of $t_{\text{min}}$ with fixed $t_{\text{max}}=32$. Middle: effective masses (blue points) from the two-point function, compared with $aM_{D_s}$ from the fit (orange band) with fitting range $15\leq t \leq 32$. Right: fitted $aM_{D_s}$ as a function of the number of states from the multi-state fit in the range $2\leq t\leq 32$.}
 \label{lqcd mass tmin}
\end{figure}

Taking the fit for $M_{D_s}$ on ensemble \texttt{f004} as an example, the left panel of Fig.~\ref{lqcd mass tmin} shows the fitted mass $aM_{D_s}$ in lattice units obtained by varying $t_{\rm min}$ for the fitting range $[t_{\text{min}}, T/2]$. Here, as $t_{\rm min}$ decreases, the contamination from excited states becomes more and more significant. It can be seen that $aM_{D_s}$ shifts upwards when $t_{\text{min}}$ is less than 14. We have chosen the fitting range as $15\leq t \leq 32$, with the corresponding $\chi^2/\text{d.o.f}=0.9$. The blue points on the middle panel of Fig.~\ref{lqcd mass tmin} represent the effective mass defined by
\begin{equation}
    aM_{\text{eff}}={\rm ArcCosh}\left[\frac{C(t+1)+C(t-1)}{2C(t)}\right],
\end{equation}
while the orange band represents $aM_{D_s}$ obtained from the single-state fit with the fitting range $15\leq t \leq 32$. It can be observed that the two are consistent within the margin of error.
The right panel of Fig.~\ref{lqcd mass tmin} shows $aM_{D_s}$ from the multi-state fits with the fitting range $2\leq t\leq 32$ as a function of the number of states $n$ in the fitting function~\cite{Lepage:2001ym}. After $n$ exceeds $4$, the fitting results stabilize and agree with that from the one-state fit.

For the other charmed mesons and charmonia, we have adopted similar procedures to determine their fitting ranges. We found that different valence quark masses do not affect very much the behavior of the effective mass with respect to the lattice time $t$. Therefore, for the same meson with different valence quark masses, we have chosen the same value of $t_{\text{min}}$. The fitted results for $aM_{D_s}$ from one-state fits on ensemble \texttt{f004}, as well as the statistical errors obtained by Jackknife analyses, are listed in Table~\ref{lqcd tab mds}. These 24 different meson masses are not the final physical results. What we want to obtain are the meson masses and decay constants at the physical mass point of valence quarks.

We also obtained the masses of pion and kaon. 
Then we use the aforementioned $m_\pi^{\rm phys}$, $(m_{ss}^{\rm phys})^2\equiv2(m_K^{\rm phys})^2-(m_\pi^{\rm phys})^2$, and $m_{J/\psi}^{\rm phys}$
as inputs to determine the physical meson masses and decay constants. From Table~\ref{lqcd tab mpk}, it can be seen that $(am_\pi)^2$ shows a linear dependence on the light quark mass, while $(am_{ss})^2$ is independent of it but shows a linear dependence on the strange quark mass, which are consistent with the expectations of chiral perturbation theory. Note that there is another method of determining the physical mass point of the strange quark, which is carried out by calculating $m_{\eta_s}$ directly, instead of $m_{ss}$.
Here $\eta_s$ is a fictitious $s\bar s$ pseudoscalar meson whose two-point functions are calculated by considering the contributions only from the QCD-connected contraction. By using $m_{\eta_s}^{\rm phys}=689.89$ MeV as an input \cite{Du:2024wtr,Borsanyi:2020mff}, we obtained $r_{\rm phys}\equiv (m_{\eta_s}^{\rm phys})^2/(m_{ss}^{\rm phys})^2=1.004(16)$ and  $f_{\eta_s}=179.6(4.1)$ MeV, which are consistent with the previous results \cite{Dowdall:2013rya,Aoki:2007xm,Davies:2009tsa}, and found that the final results of the decay constants of charmed mesons get changed by less than 0.2$\%$. We show in Table~\ref{lqcd tab metas} some fitted results of $am_{\eta_s}$ and $r$ on ensemble \texttt{f004} as examples.
\begin{table}[!htbp]
    \caption{Fitted masses of $D_s$ at various valence quark masses, with statistical uncertainties estimated by Jackknife analyses.}
    \label{lqcd tab mds}
    \centering
    \setlength{\tabcolsep}{4pt}
    \renewcommand{\arraystretch}{1.2}
\begin{tabular}{cccccccc}
\hline\hline
$am_s$     & 0.037 & 0.040 & 0.043 &  0.046&  0.049 & 0.052 & $am_c$\\
\hline
$aM_{D_s}$ & 0.7733(9)&0.7763(9)&0.7793(8)&0.7824(8)&0.7854(8)&0.7884(8) &  $0.450$\\
         & 0.8179(10) & 0.8208(9)& 0.8238(9) & 0.8267(9) & 0.8297(8)& 0.8327(8) &  $0.492$\\
         & 0.8263(10) & 0.8293(9) & 0.8322(9) & 0.8352(9) & 0.8381(9) & 0.8411(8) &  $0.500$\\
         & 0.8791(11) &0.8820(10) &0.8849(10) &0.8878(10) &0.8907(9) &0.8935(9) &  $0.550$\\
\hline\hline
\end{tabular}
\end{table}
\begin{table}[!htbp]
    \caption{Fitted results of $m_\pi^2$ and $m_{ss}^2$ at various valence quark masses, with statistical uncertainties estimated by Jackknife analyses.}
    \label{lqcd tab mpk}
    \centering
    \setlength{\tabcolsep}{3pt}
  \renewcommand{\arraystretch}{1.2}
\begin{tabular}{ccccccccc}
\hline\hline
$am_l$     & 0.0046 & 0.0585 & 0.0677&$\cdots$ &  0.0152&  0.018 & 0.024 & $am_s$\\
\hline
$(am_{\pi})^2$ & 0.0086(2)&0.0110(2)&0.0127(2)&$\cdots$&0.0282(2)&0.0333(3)&0.0442(3) &  \\
$(am_{ss})^2$      & 0.0696(10) & 0.0696(9)& 0.0696(9) &$\cdots$& 0.0697(8) & 0.0698(8)&0.0700(8) &  $0.037$\\
$(am_{ss})^2$         & 0.0751(10) & 0.0751(9)& 0.0751(9) &$\cdots$& 0.0753(8) & 0.0754(8)&0.0757(8) &  $0.040$\\
$(am_{ss})^2$         & 0.0807(10) & 0.0807(10)& 0.0807(9) &$\cdots$& 0.0809(9) & 0.0810(8)&0.0814(8) &  $0.043$\\
$(am_{ss})^2$         & 0.0862(10) & 0.0863(9)& 0.0863(9) &$\cdots$& 0.0866(9) & 0.0867(9)&0.0871(8) &  $0.046$\\
$(am_{ss})^2$         & 0.0918(11) & 0.0919(10)& 0.0919(10) &$\cdots$& 0.0922(9) & 0.0924(9)&0.0928(9) &  $0.049$\\
$(am_{ss})^2$         & 0.0974(11) & 0.0975(11)& 0.0975(10) &$\cdots$& 0.0979(9) & 0.0981(9)&0.0986(9) &  $0.052$\\
\hline\hline
\end{tabular}
\end{table}

\begin{table}[!htbp]
	\caption{Fitted results of $r\equiv m_{\eta_s}^2/(2m_K^2-m_{\pi}^2)$ and $am_{\eta_s}$ at various strange quark masses on ensemble \texttt{f004}, with statistical uncertainties estimated by Jackknife analyses.}
	\label{lqcd tab metas}
	\centering
	\setlength{\tabcolsep}{3pt}
	\renewcommand{\arraystretch}{1.2}
	\begin{tabular}{ccccccc}
		\hline\hline
		$am_s$     & 0.037 & 0.040 & 0.043& 0.046 &  0.049&  0.052 \\
		\hline
		$r$ & 0.996(8) & 0.999(8) & 1.002(8) &   1.005(8) &  1.008(7) &  1.011(7) \\ 
		$am_{\eta_s}$ & 0.2632(7)  & 0.2739(7)  & 0.2844(7)  & 0.2945(7)  & 0.3043(7)  & 0.3140(7) \\
		\hline\hline
	\end{tabular}
\end{table}

\begin{figure}[!htbp]
    \centering
   \includegraphics[width=0.42 \columnwidth]{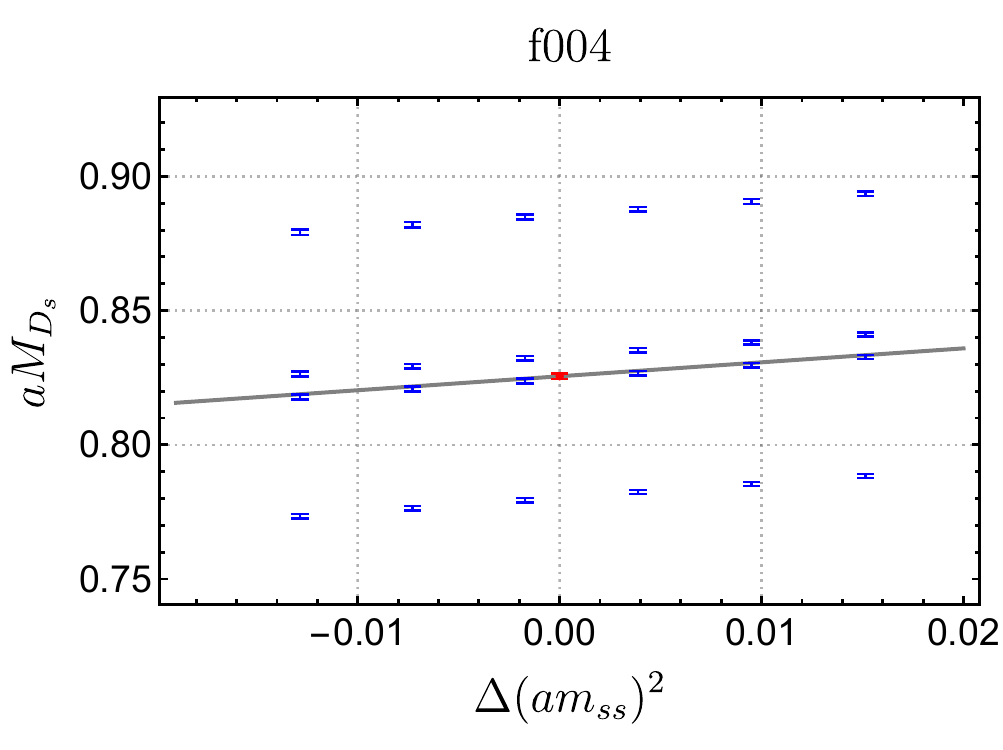}
   \includegraphics[width=0.42 \columnwidth]{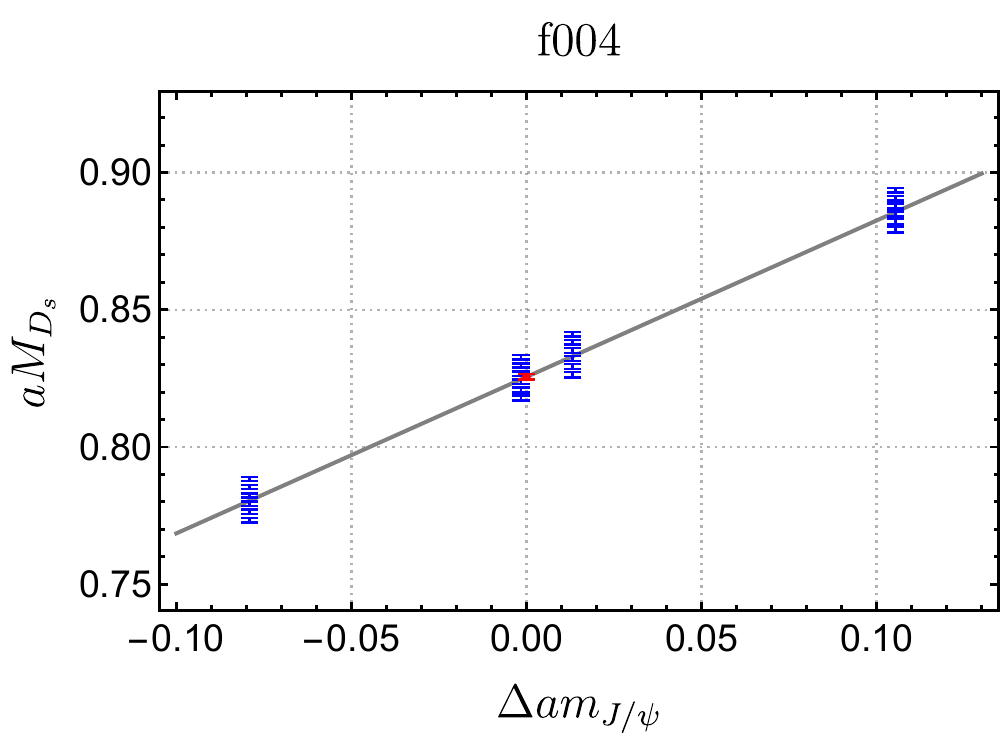}
   \includegraphics[width=0.42 \columnwidth]{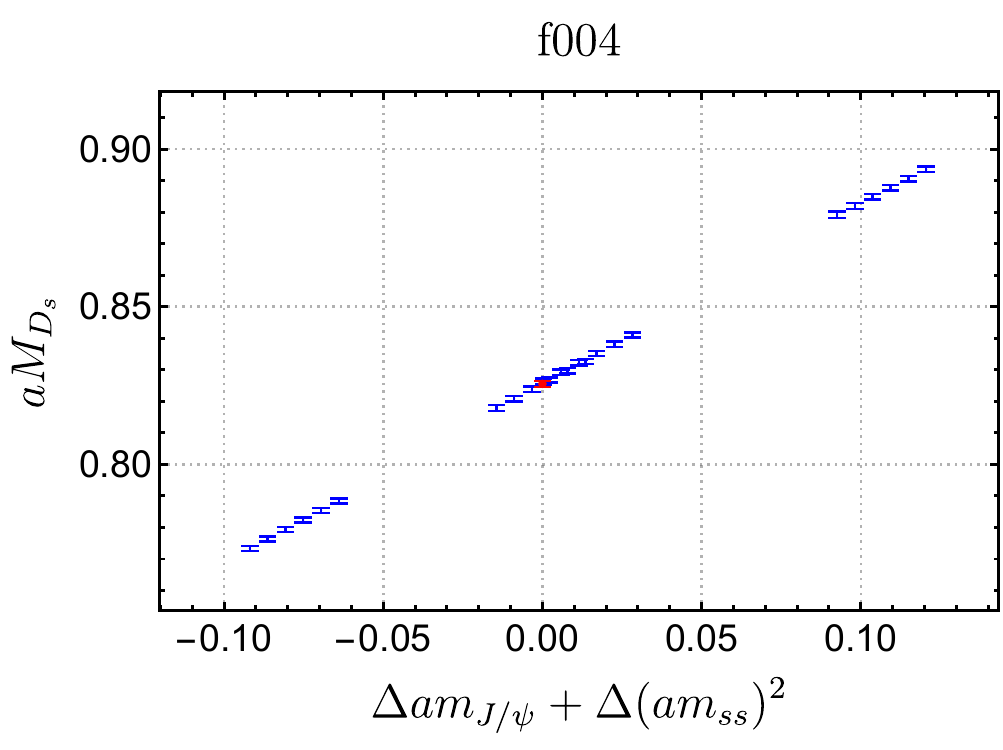}
    \caption{ Distribution of $aM_{D_s}$ (blue dots) with respect to $\Delta (am_{ss})^2$ and/or $\Delta am_{J/\psi}$, and linear interpolation of $aM_{D_s}$ (gray lines) to the physical point (red dot).}
 \label{lqcd fig explor}
\end{figure}

After obtaining the meson masses and decay constants at different valence quark masses on ensemble \texttt{f004}, we used linear interpolation (extrapolation) to obtain the results at the physical mass point. The Particle Data Group~\cite{ParticleDataGroup:2022pth} and Fermilab Lattice and MILC Collaborations~\cite{Bazavov:2017lyh} give $$m_\pi^{\rm phys}=134.98\ \text{MeV},\quad m_K^{\rm phys}=494.49\ \text{MeV},\quad m_{J/\psi}^{\rm phys}=3.0969\ \text{GeV}.$$ Combining the above with the inverse lattice spacing $a^{-1}=2.383(9)\,$GeV, we have $$(am_\pi)^2_{\rm phys}=0.00321(2)\,,\ (am_{ss})^2_{\rm phys}=0.0829(7)\,,\ (am_{J/\psi})_{\rm phys}=1.2996(49)\,.$$ Here, we regard the error from hadron masses as tiny and negligible, compared with that from our lattice spacing. The form of the linear interpolation function is as follows,
\begin{equation}
 aM_{D_s}-aM_{D_s}^{\rm fit}=b_1\Delta am_{J/\psi}+b_2\Delta (am_{ss})^2,
    \label{lqcd:eq:extrapolate}
\end{equation}
 with $\Delta am_{J/\psi}=am_{J/\psi}-(am_{J/\psi})_{\rm phys}$ and $\Delta (am_{ss})^2=(am_{ss})^2-(am_{ss})^2_{\rm phys}$ as inputs, and $b_1$, $b_2$ and $aM_{D_s}^{\rm fit}$ as fitting parameters. On ensemble \texttt{f004}, we can then obtain
\begin{equation}
    aM_{D_s}^{\rm fit}=0.8258(10),
\end{equation}
where the statistical error also comes from Jackknife analyses, taking into account the correlation among different valence quark masses. An illustrative plot of the interpolation for the 24 data points of $aM_{D_s}$ in Table~\ref{lqcd tab mds} is shown in Fig.~\ref{lqcd fig explor}. The four rows of blue points in the figure correspond to the variation of $aM_{D_s}$ with respect to $(am_{ss})^2$ for the four different charm quark masses. The red point represents the result of the interpolation at the physical mass point. It can be seen that Eq.~(\ref{lqcd:eq:extrapolate}) describes the data well, and the physical mass point of the valence charm quark is around $am_c=0.492$.

The computed $D^{(*)}$ masses on configuration \texttt{f004} are listed in Table~\ref{lqcd tab md}. Similarly, we used linear interpolation and extrapolation methods to calculate the $D^{(*)}$ mass at the physical mass point. The fitting function for the interpolation and extrapolation is given by
\begin{equation}
 aM_{D^{(*)}}-aM_{D^{(*)}}^{\rm fit}=b_1\Delta am_{J/\psi}+b_3\Delta (am_{\pi})^2.
    \label{lqcd:eq:extrapolate2}
\end{equation}
Here the mass of $D^{(*)}$ depends on the light quark mass, so we included $b_3\Delta (am_{\pi})^2$ in Eq.~(\ref{lqcd:eq:extrapolate2}), where $\Delta (am_{\pi})^2=(am_\pi)^2-(am_\pi)^2_{\rm phys}$. The fitting results are
\begin{equation}
     aM_{D}^{\rm fit}=0.7837(30),\qquad  aM_{D^*}^{\rm fit}=0.8522(57).
\end{equation}
The relative error of $aM_{D^{(*)}}^{\rm fit}$ is larger than that of $aM_{D_s^*}$ because our valence pion masses ranging from 220 to 500 MeV are larger than the physical one, and thus an extrapolation in $(am_\pi)^2$ is needed.
\begin{table}[!htbp]
    \caption{ $M_{D^{(*)}}$ with statistical uncertainties estimated by Jackknife analyses..}
    \label{lqcd tab md}
    \centering
    \setlength{\tabcolsep}{4pt}
   \renewcommand{\arraystretch}{1.2}
\begin{tabular}{ccccccccc}
\hline\hline
$am_l$     & 0.0046 & 0.0585 & 0.0677&$\cdots$ &  0.0152&  0.018 & 0.024 & $am_c$ \\
\hline
$aM_{D}$ & 0.7417(28)&0.7423(24)&0.7429(22)&$\cdots$&0.7507(14)&0.7535(13)&0.7595(11) &  
$0.450$ \\
       & 0.7867(31)&0.7873(27)&0.7879(25)&$\cdots$&0.7955(15)&0.7982(14)&0.8042(12) &   $0.492$\\
       & 0.7953(31)&0.7958(27)&0.7964(25)&$\cdots$&0.8040(16)&0.8067(14)&0.8126(12) &     $0.500$\\
      & 0.8484(35)&0.8489(30)&0.8495(28)&$\cdots$&0.8570(17)&0.8596(16)&0.8655(13) &    $0.550$\\
\hline
$aM_{D^{*}}$ 
       & 0.8129(56) & 0.8140(49) & 0.8147(46)&$\cdots$&0.8211(28) & 0.8236(25) & 0.8292(21) &   $0.450$\\
       &  0.8541(58) & 0.8553(51) & 0.8560(47) &$\cdots$&0.8625(29) & 0.8650(26) & 0.8706(22)&     $0.492$\\
      &  0.8620(59) & 0.8631(52) & 0.8639(48)&$\cdots$&0.8704(29) & 0.8729(26) & 0.8785(22) &    $0.500$\\
       &0.9112(62) & 0.9125(54) & 0.9133(50) & $\cdots$ & 0.9199(31) & 0.9223 (28)& 0.9279(23) &$0.550$\\
\hline\hline
\end{tabular}
\end{table}

We repeated the previous steps on ensemble \texttt{f006} and \texttt{f008}, and combined the results at the physical mass points with those from \texttt{f004}, all of which are listed in Table~\ref{lqcd tab 3m}. For the mass of $\eta_c$ an interpolation only on $am_{J/\psi}$ is performed.

The sea quark masses $m_l^{\rm sea}$ for none of the three ensembles are at the physical mass point. Therefore, we need to perform another round of linear extrapolation to obtain the final physical results (taking $aM_{D_s}$ as an example),
\begin{equation}
   aM_{D_s}^{\rm fit}-aM_{D_s}^{\rm phys}=b_5\left[(am_\pi^{\rm sea})^2-(am_\pi)^2_{\rm phys} \right],
    \label{lqcd uni}
\end{equation}
where $aM_{D_s}^{\rm fit}$ and $(am_\pi^{\rm sea})^2$ are taken from Table~\ref{lqcd tab 3m}, 
$m_\pi^{\rm sea}$ refers to the pion masses corresponding to the light sea quark masses~\cite{RBC:2010qam}, and $aM_{D_s}^{\rm phys}$ is a fitting parameter. In Eq.~(\ref{lqcd uni}), $aM_{D_s}$ can be replaced by other observables.
\begin{table}[!htbp]
    \caption{Fitted masses of charmed mesons and charmonia on ensembles \texttt{f004}, \texttt{f006} and \texttt{f008} at physical mass point of valence quarks, with $m_\pi^{\rm sea}$ the pion masses corresponding to the light sea quark masses~\cite{RBC:2010qam}.}
    \label{lqcd tab 3m}
    \centering
   \setlength{\tabcolsep}{4pt}
  \renewcommand{\arraystretch}{1.3}
\begin{tabular}{ccccccc}
\hline\hline
  Label &   $am_\pi^{\rm sea}$ & $aM_D^{\rm fit}$ & $aM_{D^*}^{\rm fit}$ & $aM_{D_s}^{\rm fit}$  & $aM_{D_s^*}^{\rm fit}$  & $aM_{\eta_c}^{\rm fit}$   
  \\
\hline
  \texttt{f004} &  0.1269(4) &0.7837(30)  &0.8521(57)  &0.8258(10)  &0.8914(13) & 1.24967(59)\\
 \texttt{f006} &   0.1512(3) & 0.7836(27) &0.8644(34) &0.8261(9) &0.8951(12) & 1.24796(59)  \\
 \texttt{f008} &  0.1727(4)  & 0.7843(25) &0.8684(32) &0.8264(8) &0.8964(18) &1.24884(52)\\
\hline\hline
\end{tabular}
\end{table}

\begin{figure}[!htbp]
    \centering
   \includegraphics[width=0.46 \columnwidth]{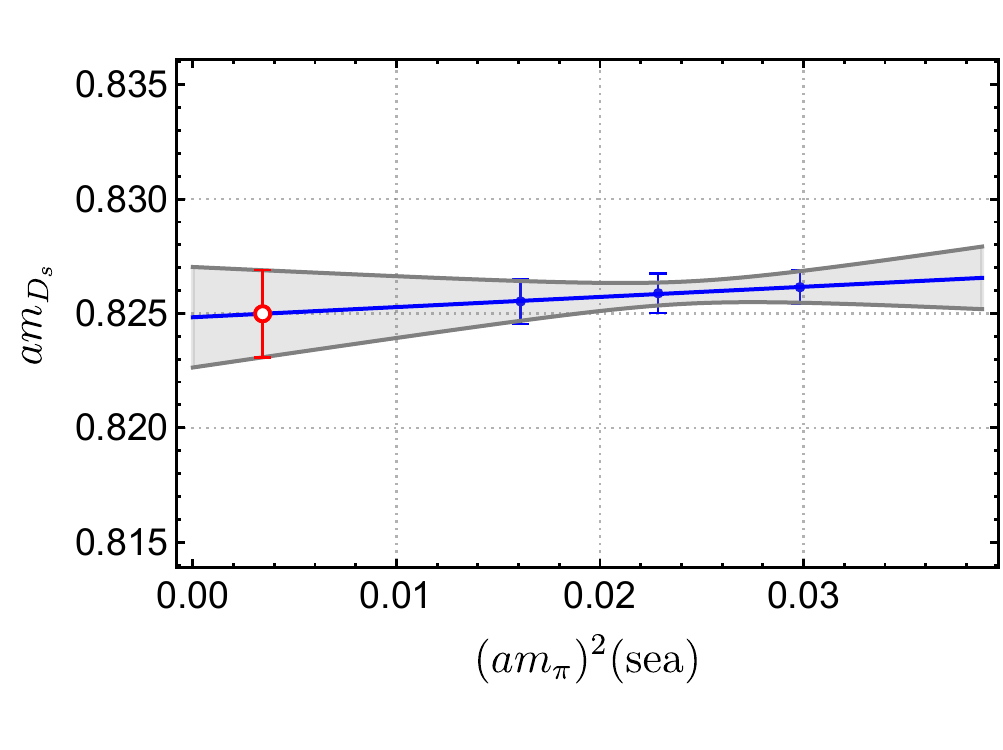}
    \caption{Extrapolations of $aM_{D_s}^{\rm fit}$ (blue) to the physical mass point of light sea quarks (red) with uncertainties (gray band). The horizontal coordinates are the pion masses squared $(am_\pi^{\rm sea})^2$ in lattice units corresponding to the sea quark masses on ensembles \texttt{f004}, \texttt{f006} and \texttt{f008}.}
 \label{waitui}
\end{figure}

Fig.~\ref{waitui} shows the extrapolation of $aM_{D_s}$ with respect to the sea quark masses by using Eq.~(\ref{lqcd uni}). Meson masses do not get corrections from renormalization, so we can multiply the results in lattice units by the inverse lattice spacing $a^{-1}=2.383(9)\,$GeV to obtain the physical results,
\begin{equation}
\begin{aligned}
    &M_D^{\text{phys}}=1.866(14)(7)\,\text{GeV},\quad M_{D^*}^{\text{phys}}=2.007(24)(8)\,\text{GeV},\\
  &  M_{D_s}^{\text{phys}}=1.9666(46)(75)\,\text{GeV}, \quad
    M_{D_s^*}^{\text{phys}}=2.1133(71)(80)\,\text{GeV}, \\ &\qquad\qquad\qquad M_{\eta_c}^{\text{phys}}=2.978(3)(12)\,\text{GeV}.
\end{aligned}
\end{equation}
Here the first error is statistical and the second one is from the uncertainty of the lattice spacing. The statistical errors for $D^{(*)}$ are much larger than those for $D^{(*)}_s$ because going to the physical mass point of the light valence quark is implemented by an extrapolation rather than interpolation.

\begin{figure}[!htbp]
    \centering
   \includegraphics[width=0.46 \columnwidth]{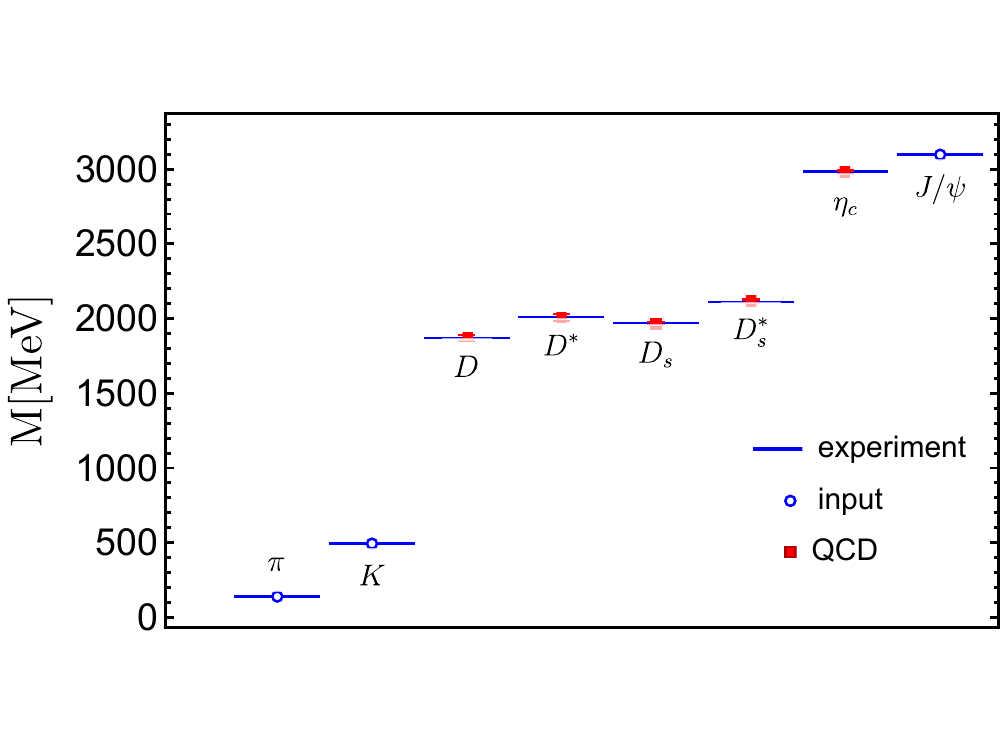}
 \includegraphics[width=0.46 \columnwidth]{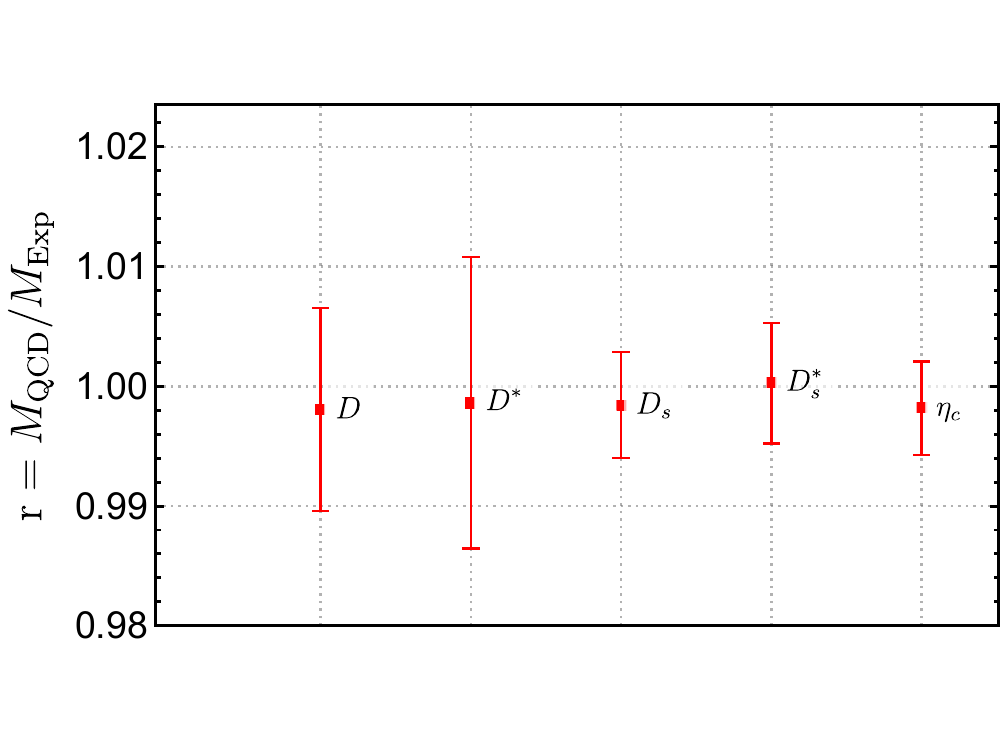}
    \caption{Comparisons of our meson masses with experimental values. The experimentally measured masses $m_\pi^{\rm expt}$, $m_K^{\rm expt}$, and $m_{J/\psi}^{\rm expt}$ are used as inputs. The right panel shows the ratios of our meson masses to their experiment values. Our results are in good agreement with experiment measurements.}
 \label{fig:exp-LQCD}
\end{figure}

As shown in Fig.~\ref{fig:exp-LQCD}, our meson masses are consistent with the values in Particle Data Group within $1\sigma$ deviation:$$M_{D^{\pm}}^{\rm expt}=1.86965(5)\,\text{GeV},\quad M_{D^{*\pm}}^{\rm expt}=2.01026(5)\,\text{GeV}$$ $$M_{D_s}^{\rm expt}=1.96835(7)\,\text{GeV},\quad M_{D_s^{*}}^{\rm expt}=2.1122(4)\,\text{GeV},\quad M_{\eta_c}^{\rm expt}=2.9839(5)\,\text{GeV}.$$
The right panel of Fig.~\ref{fig:exp-LQCD} shows the ratios of our meson masses to their experiment values, which are consistent with one.
Compared to the results at a coarse lattice spacing $a^{-1}=1.730(4)\,$GeV~\cite{Chen:2020qma}: $$M_{D^{\pm}}^{\rm latt}=1.873(5)\,\text{GeV},\quad M_{D^{*\pm}}^{\rm latt}=2.026(5)\,\text{GeV},\quad M_{D_s^{*}}^{\rm latt}=2.116(6)\,\text{GeV},$$ our current results are also consistent within $1\sigma$ deviation. The largest difference in the center values between the current and previous results is in the mass of $D^*$, which is about 1$\%$. Therefore, we estimate the discretization error of our masses of charmed mesons and charmonia to be approximately $1\%$. This confirms the estimation of the discretization error given in the previous work~\cite{Chen:2020qma}.

\subsection{Decay constants}
\label{sec:fx}
Before analyzing the decay constants, we first provide the renormalization constants for the vector and tensor currents. As mentioned before, we use chiral lattice fermions in this work. Thus, the decay constants $f_P$ obtained from the two-point functions of pseudoscalar operators do not get corrections from renormalization. The tensor current is renormalized by $Z_T$ in the the $\overline{\text{MS}}$ scheme at the scale of $2\,$GeV, and the renormalization constant for the vector current is the same as that of the axial-vector one, viz. $Z_V=Z_A$, of which the numerical results are copied in Eq.~(\ref{eq:Zfactors}) from Ref.~\cite{Bi:2023pnf}.

Following the fitting procedure introduced in the previous section, we use single-state fits, Eq.~(\ref{eq:ct-func0}), for two-point correction functions $C(t)$.
The decay constant can be obtained from the amplitude $A_0$. For pseudoscalar mesons, the decay constant is given by
\begin{equation}
f_P=\frac{m_{q_1}+m_{q_2}}{(m_0)^{3/2}}\sqrt{2A_0},
    \label{eq pcsa}
\end{equation}
where $q_1$ and $q_2$ represent the quark components in the pseudoscalar operator. For vector mesons, the decay constant (before renormalization) is given by
\begin{equation}
f_V=\frac{1}{(m_0)^{1/2}}\sqrt{2A_0}
\label{eq vc}
\end{equation}
from fits to the correlation functions with vector current inserted.

In addition to calculating the decay constants themselves, we also performed joint fits of two-point functions 
and obtained the ratio of decay constants $f_V/f_P$, as well as $f^T_V/f_V$. 

\begin{figure}[!htbp]
    \centering
   \includegraphics[width=0.6 \columnwidth]{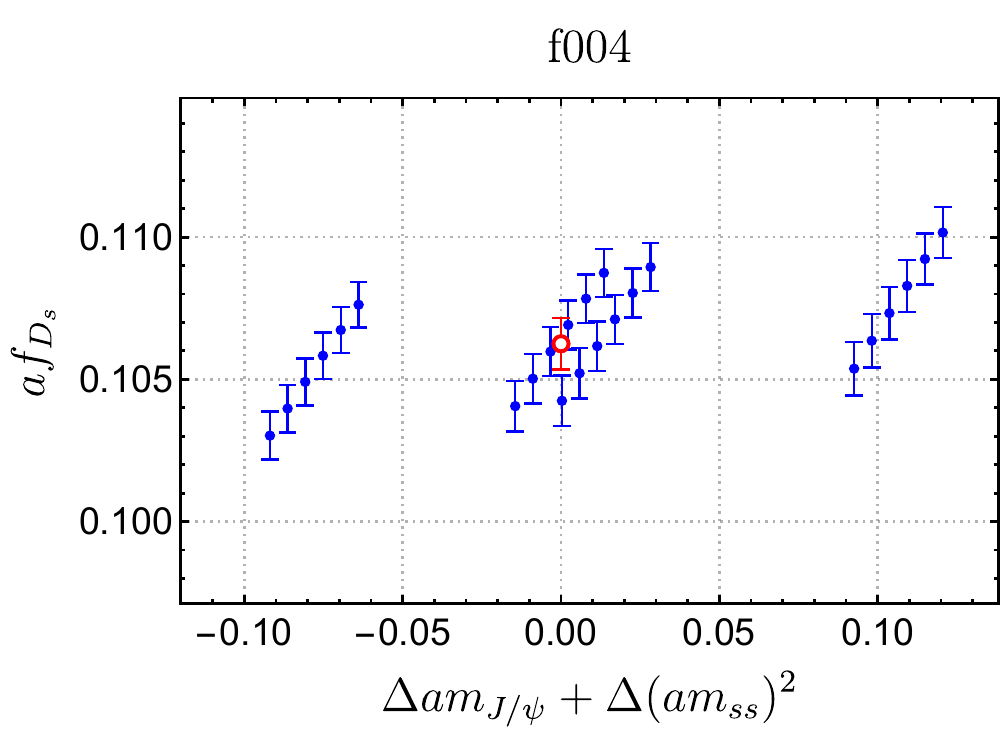}
    \caption{ Distribution of $af_{D_s}$ (blue dots) with respect to $\Delta am_{J/\psi}+\Delta (am_{ss})^2$, with linear interpolation to the physical mass point (red dot).}
 \label{lqcd fig decay}
\end{figure}

The subsequent steps for going to the physical mass point are similar to those in the previous section. First, we collected decay constants and their ratios on different ensembles at different valence quark masses. Some of the results are listed in Table~\ref{lqcd tab decay}.
\begin{table}[!htbp]
    \caption{Decay constants of $D_s$ extracted from $\chi^2$-fit on ensemble \texttt{f004}, with statistical uncertainties estimated by Jackknife analyses.}
    \label{lqcd tab decay}
    \centering
   \setlength{\tabcolsep}{4pt}
   \renewcommand{\arraystretch}{1.2}
\begin{tabular}{cccccccc}
\hline\hline
$am_s$     & 0.037 & 0.040 & 0.043 &  0.046&  0.049 & 0.052 & $am_c$ \\
\hline
$af_{D_s}$  
 &0.1030(9) & 0.1040(9) & 0.1050 (9)& 0.1059 (9)& 0.1067 (9)& 0.1076 (9)&$0.45$\\
 &0.1041(9) & 0.1050(9) & 0.1060(9) & 0.1069(9) & 0.1078 (9)& 0.1087 (9)&$0.492$\\
 &0.1042(9) & 0.1052(9) & 0.1062(9) & 0.1071(9) & 0.1080(9) & 0.1089(9) &$0.50$\\
 &0.1054(10) & 0.1064(10) & 0.1073(10) & 0.1083(10) & 0.1092(9) & 0.1102(9) &$0.55$\\
\hline\hline
\end{tabular}
\end{table}
We then performed a linear interpolation (extrapolation) of the results on each ensemble. Taking $f_{D_s}$ on \texttt{f004} as an example, the interpolation function takes the form similar to Eq.~(\ref{lqcd:eq:extrapolate})
\begin{equation}
 af_{D_s}-(af_{D_s})_{\text{fit}}=b'_1\Delta am_{J/\psi}+b'_2\Delta (am_{ss})^2.
\end{equation}
The corresponding figure is shown in Fig.~\ref{lqcd fig decay},
where the abscissa values are of $\Delta am_{J/\psi}+\Delta (am_{ss})^2$. It is seen in the figure that the linear fitting function describes our data well. Some decay constants and ratios fitted at the physical mass point are collected in Table~\ref{lqcd tab 3f}.
\begin{table}[!htbp]
    \caption{Decay constants before renormalization and their ratios on ensembles \texttt{f004}, \texttt{f006} and \texttt{f008} of charmed mesons and charmonia fitted at the physical mass point of valence quarks, with $m_\pi^{\text{sea}}$ the pion masses corresponding to the sea quark masses~\cite{RBC:2010qam}.}
    \label{lqcd tab 3f}
    \centering
   \setlength{\tabcolsep}{4pt}
  \renewcommand{\arraystretch}{1.3}
\begin{tabular}{ccccccc}
\hline\hline
  Label &   $am_\pi^{\text{sea}}$ & $(af_D)_{\text{fit}}$ & $(af_{D^*})_{\text{fit}}$ & $(af_{D^*}^T)_{\text{fit}}$  & $(f_{D^*}/f_D)_{\text{fit}}$  & $f^T_{D^*}/f_{D^*}(\text{fit})$
  \\
\hline
  \texttt{f004} &  0.1269(4) &0.0907(18)  &0.0944(38)  &0.0797(24)  &1.041(41) & 0.845(28)\\
 \texttt{f006} &   0.1512(3) & 0.0920(18) &0.1049(22) &0.0870(25) &1.139(26) & 0.829(17)  \\
 \texttt{f008} &  0.1727(4)  & 0.0915(18) &0.1079(19) &0.0918(15) &1.179(29) &0.850(9)\\
\hline\hline
  Label &   $am_\pi^{\text{sea}}$ & $(af_{D_s})_{\text{fit}}$ & $(af_{D_s^*})_{\text{fit}}$ & $(af_{D_s^*}^T)_{\text{fit}}$  & $(f_{D_s^*}/f_{D_s})_{\text{fit}}$  & $(f^T_{D_s^*}/f_{D_s^*})_{\text{fit}}$
  \\
\hline
  \texttt{f004} &  0.1269(4) &0.1063(10)  &0.1103(11)  &0.0937(10)  &1.038(12) & 0.8492(43)\\
 \texttt{f006} &   0.1512(3) & 0.1061(9) &0.1126(12) &0.0963(11) &1.061(14) & 0.8550(63)  \\
 \texttt{f008} &  0.1727(4)  & 0.1054(8) &0.1131(16) &0.0962(14) &1.073(17) &0.8505(71)\\
\hline\hline
  Label &   $am_\pi^{\text{sea}}$ & $(af_{\eta_c})_{\text{fit}}$ & $(af_{J/\psi})_{\text{fit}}$ & $(af_{J/\psi}^T)_{\text{fit}}$  & $(f_{J/\psi}/f_{\eta_c})_{\text{fit}}$  & $(f^T_{J/\psi}/f_{J/\psi})_{\text{fit}}$
  \\
\hline
  \texttt{f004} &  0.1269(4) &0.1748(12)  &0.1704(11)  &0.1527(11)  &0.9754(70) & 0.8960(14)\\
 \texttt{f006} &   0.1512(3) & 0.1753(9) &0.1760(10) &0.1573(10) &1.004(66) & 0.8935(14)  \\
 \texttt{f008} &  0.1727(4)  & 0.1731(8) &0.1699(11) &0.1521(10) &0.9818(57) & 0.8950(16)\\
\hline\hline

\end{tabular}
\end{table}

\begin{figure}[!htbp]
    \centering
  
   \includegraphics[width=0.46 \columnwidth]{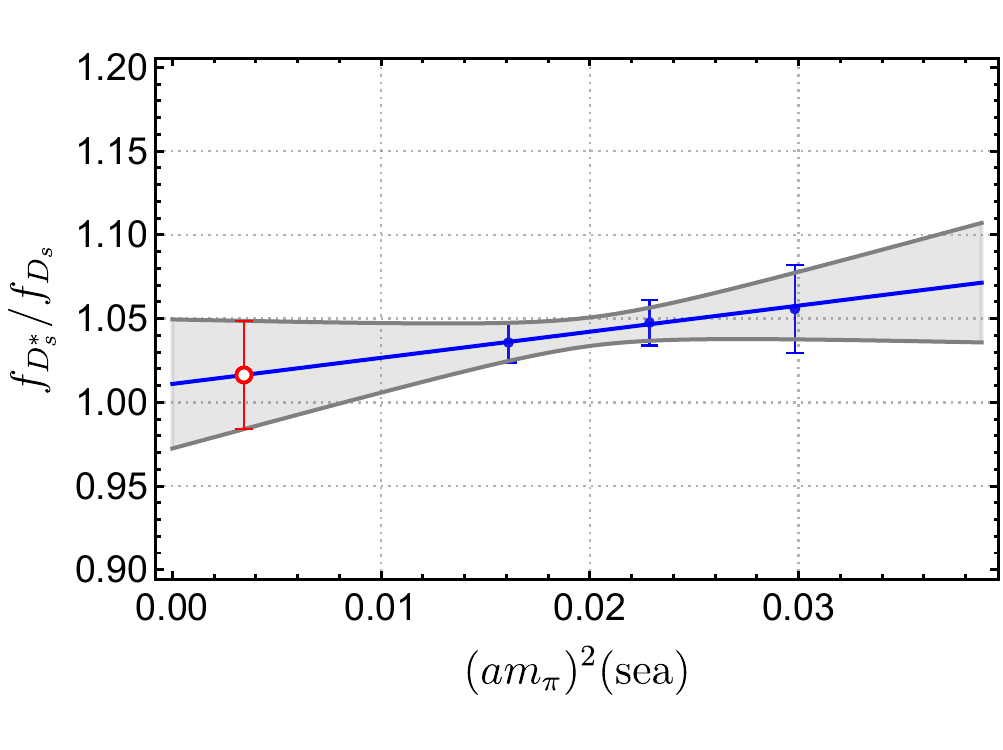}
   \includegraphics[width=0.46 \columnwidth]{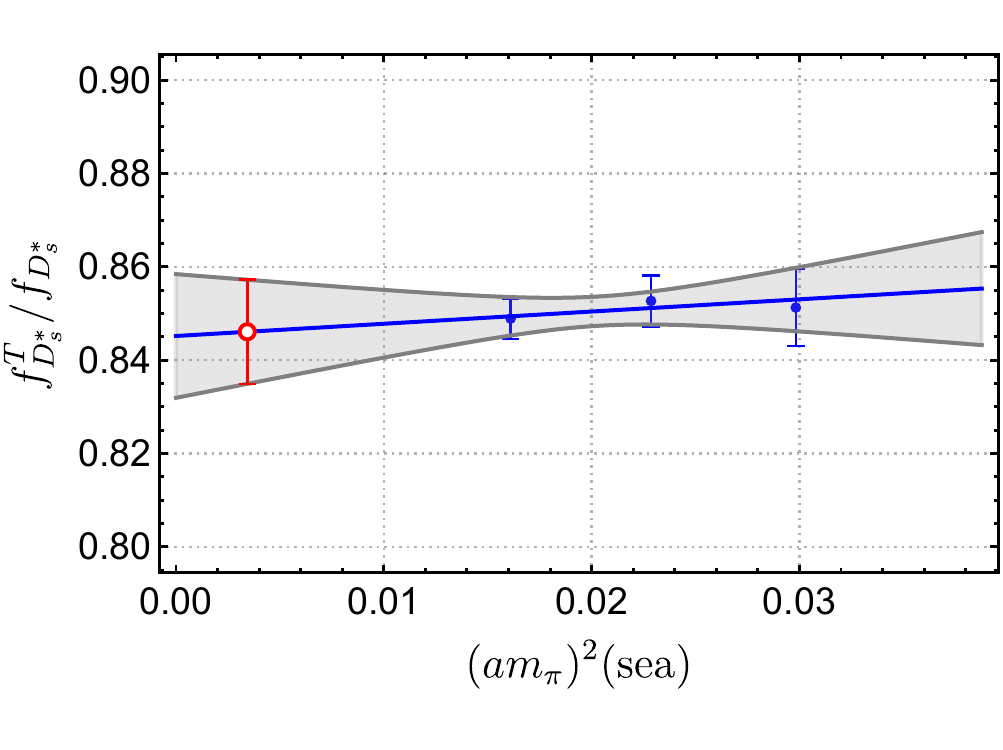}
   \includegraphics[width=0.46 \columnwidth]{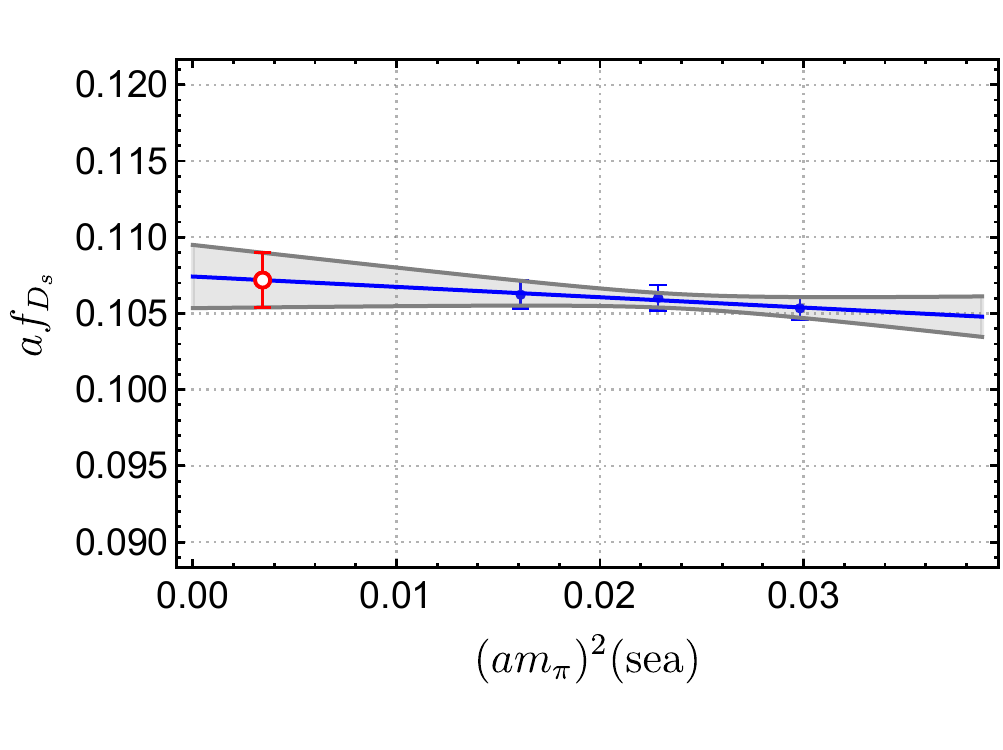}
    \caption{Examples of extrapolations of 
    decay constants or ratios (blue) to the physical mass point (red) of light sea quarks with uncertainties (gray band). The abscissa values are of pion masses squared $(am_\pi^{\rm sea})^2$ in lattice units corresponding to the light sea quark masses on ensembles \texttt{f004}, \texttt{f006} and \texttt{f008}.}
 \label{lqcd fig explor2}
\end{figure}

Finally, we performed a linear extrapolation for the sea quark mass $m_l^{\rm sea}$ to obtain the decay constants at the physical sea quark mass point, as shown in Fig.~\ref{lqcd fig explor2}.
We converted the decay constants in lattice units to physical units and multiplied them by the appropriate renormalization constants. The physical results are as follows (in units of MeV),
\begin{equation}
\begin{aligned}
    &  f_{D}=215.3(9.1)(0.8),\quad f_{D^*}=223.7(16.2)(0.9),\quad f_{D^*}^T=190.8(12.6)(1.9)(0.8),\\
    & \, f_{D_s}=255.7(4.3)(1.0),\quad f_{D_s^*}=276.8(6.4)(1.1),\quad f_{D_s^*}^T=251.9(6.0)(2.4)(1.0).
\end{aligned}
\label{eq:fD}
\end{equation}
Here, for each of the decay constants, the first error includes statistical and extrapolation/interpolation uncertainties, the last error is from the uncertainty of the lattice spacing, and the error in the middle, if there is one, is due to the uncertainty in the renormalization constant.
It is seen in Eq.~(\ref{eq:Zfactors}) that the uncertainty of $Z_V(=Z_A)$ can be regarded as negligible. The uncertainty of $Z_T$ is around 1\% and is much smaller than the first error.

For chiral fermions as used in this work the discretization effects are proportional to the squared lattice spacing $a^2$.
With three or more lattice spacings we would be able to do linear fittings in $a^2$ and obtain results in the continuum limit. The lattice spacing squared corresponding to our current work ($a^2\approx 0.007$ fm$^2$) happens to lie almost at the middle point between the continuum limit and the previous work~\cite{Chen:2020qma} ($a^2\approx 0.013$ fm$^2$). A simple linear extrapolation in $a^2$ using the results at the two lattice spacings has no degree of freedom and will shift our current result by an amount almost equal to the difference between the two lattice spacings.
Therefore, we think the discretization errors can be estimated by simply comparing the results of these two works. For $f_{D_s^{(*)}}$, our results 
are consistent within $1\sigma$ deviation with the previous results 
, respectively, where $f_{D_s}=249(5)\,$MeV and $f_{D_s^*}=274(5)\,$MeV~\cite{Chen:2020qma} and we have dropped the discretization error assigned to the previous work. 
For $f_{D^{(*)}}$, our results are also consistent within $1\sigma$ deviation with the previous results, respectively, where $f_{D}=213(2)\,$MeV and $f_{D^{*}}=234(3)\,$MeV. The differences in the central values are approximately from 1\% to $4\%$. Therefore, we take an average value $3\%$ as the estimate of the discretization errors in our decay constants and get (in units of MeV),
\begin{equation}
\begin{aligned}
    &  f_{D}=215.3(9.1)(6.5),\quad f_{D^*}=223.7(16.3)(6.8),\quad f_{D^*}^T=190.8(12.8)(5.8),\\
    & \, f_{D_s}=255.7(4.4)(7.7),\quad f_{D_s^*}=276.8(6.5)(8.3),\quad f_{D_s^*}^T=251.9(6.6)(7.6),
\end{aligned}
\end{equation}
where the first error is the square root of the quadratic sum of the errors in Eq.~(\ref{eq:fD}), and the second error is the 3\% discretization error, for each of the decay constants. 

The ratios of decay constants of charmed mesons are collected in Table~\ref{final results ratio} and Table~\ref{final results ratio2}, which are in agreement with those obtained in our previous work on a coarser lattice. The first error in the table takes into account the contributions from statistics, interpolation/extrapolation and renormalization. The second error is our estimate of the discretization error from the difference in the center values of this work and the previous work.
The ratios involving $D^{(*)}$ have larger errors than those from our previous work because of the extrapolation with respect to the light quark mass (both valence and sea). The result of $f_{D_s^*}^T/f_{D_s^*}$ from this work has a smaller error because the uncertainty from the renormalization factor $Z_T^{\msbar}$(2 GeV)$/Z_A$ is now smaller.

In addition, we also calculated the decay constants and their ratios for $\eta_c$ and $J/\psi$. We used the renormalization constants as shown in Eq.~(\ref{eq:Zfactors}). The decay constants themselves are 6\%-9\% higher than the values in the continuum limit obtained by other lattice groups, such as the HPQCD collaboration~\cite{Hatton:2020qhk}. We think this is due to large discretization effects in decay constants of charmonia, which are composed of two heavy quarks. In the work of HPQCD~\cite{Hatton:2020qhk}, sizable discretization effects were also spotted in those decay constants at non-zero lattice spacings.

We observed smaller discretization effects in ratios of decay constants for charmonia. The physical results are
\begin{equation}
    f_{J/\psi}/f_{\eta_c}=1.060(15),\quad  f^T_{J/\psi}/ f_{J/\psi}=0.961(10).
    \label{eq:ratio-ccbar}
\end{equation}
Here the error includes statistical uncertainty and the one in renormalization constants.
Combining these values with the experimental measured $f_{J/\psi}^{\rm expt}=407(4)\,$MeV, which is obtained from its pure leptonic decay width $\Gamma(J/\psi\to e^+e^-)$~\cite{Hatton:2020qhk}, we can derive $f_{\eta_c}=383.8(6.7)\,$MeV and $f^T_{J/\psi}=391.1(5.4)\,$MeV. These results are consistent within $1.5\sigma$ deviation with the results $f_{\eta_c}=394.7(2.4)\,$MeV~\cite{Davies:2010ip}, $f^T_{J/\psi}=392.7(2.7)\,$MeV and $f^T_{J/\psi}/ f_{J/\psi}=0.9569(52)$~\cite{Hatton:2020vzp} from the HPQCD Collaboration obtained in the continuum limit. The better agreement in the ratio means the discretization effects are shrunk by the cancellation between the numerator and the denominator.
Because we do not have results of decay constants of charmonia from the previous work to compare, we do not give discretization errors for those constants and ratios in this work.
\begin{table}[!htbp]
    \caption{Ratios of decay constants. The discretization errors of the results from~\cite{Chen:2020qma} were estimated to be 2\%.}
    \label{final results ratio}
    \centering
   \setlength{\tabcolsep}{4pt}
  \renewcommand{\arraystretch}{1.3}
\begin{tabular}{cccccccc}
 \cline{1-4} \cline{6-8}
 [This work]   &   $D^*/D$ & $D_s^*/D_s$ & $J/\psi/\eta_c$ & $\qquad$& \cite{Chen:2020qma}   &   $D^*/D$ & $D_s^*/D_s $ 
  \\
 \cline{1-4}\cline{6-8}
 $f_V/ f_{P}$ &  1.045(83)(55) & 1.097(30)(3)  & 1.060(15) & $\qquad$& $f_V/ f_{P}$ &1.10(2)(2) & 1.10(3)(2)   \\
 $f_V^T/f_V$ &  0.872(47)(38) & 0.909(14)(11) &0.961(10)   & $\qquad$&  $f_V^T/f_V$ &0.91(3)(2) &0.92(3)(2)  \\
 \cline{1-4}\cline{6-8}
\end{tabular}
\end{table}
\begin{table}[!htbp]
    \caption{Ratios of decay constants for SU(3) flavor symmetry breaking effects.}
    \label{final results ratio2}
    \centering
   \setlength{\tabcolsep}{4pt}
  \renewcommand{\arraystretch}{1.3}
\begin{tabular}{ccccccc}
\hline
    &  [This work]  & \cite{Chen:2020qma}
  \\
\hline
 $f_{D_s}/ f_D$  &  1.185(45)(23) &  1.163(14)(23)   \\
 $f_{D_s^*} / f_{D^*}$ &  1.231(73)(61) & 1.17(2)(2)\\
\hline

\end{tabular}
\end{table}

\section{Summary}
\label{sec:summary}
In the study of heavy flavor physics, decay constants of mesons are fundamental and important quantities. They are essential input parameters of theoretical calculations about semileptonic decays, such as  $B_{(c)}\to D,\eta_c$, as well as pure or non-leptonic decays of $D$ mesons, by using either the QCD-based factorization theory or phenomenological models such as topological diagram methods. Such decay constants are also crucial for extraction of CKM matrix elements and precise tests of the Standard Model. Theoretical calculations of decay constants are essential and lattice QCD provides a systematic way to improve their precision. Depending on only fundamental parameters of QCD without any modelling assumptions, lattice QCD is considered the best method of computing decay constants.

In this work, we have computed decay constants and their ratios for $D$, $D^*$, $D_s$, $D_s^*$, $\eta_c$ and $J/\psi$, using $2+1$-flavor configurations. We have also calculated the decay constants with tensor currents of vector mesons and provided estimates of discretization errors. Our final results for the decay constants adding up all errors are given in Tab.~\ref{final results decay constant} (in units of MeV), of which the ratios are given in Tabs.~\ref{final results ratio} and~\ref{final results ratio2} for charmed mesons, and in Eq.~(\ref{eq:ratio-ccbar}) for charmonia. The ratios in Tabs.~\ref{final results ratio} and~\ref{final results ratio2} reflect the magnitudes of heavy quark symmetry breaking and SU(3) flavor symmetry breaking, respectively. The precision of $f_{D_s^*}^T/f_{D_s^*}$ is improved compared with that from our previous work~\cite{Chen:2020qma}. We compare $f_{D^*_{(s)}}$ from this work and other lattice QCD calculations in Tab.~\ref{comparison}. 
\begin{table}[!htbp]
    \caption{Decay constants in units of MeV, among which the  $f_{\eta_c}$ and $f_{J/\psi}^T$ require $f_{J/\psi}$ as an additional input. The ones with tensor currents are renormalized in the $\msbar$ scheme at the scale of 2 GeV.}
    \label{final results decay constant}
    \centering
   \setlength{\tabcolsep}{4pt}
  \renewcommand{\arraystretch}{1.3}
\begin{tabular}{cccccccc}
\cline{1-5}\cline{7-8}
    &   $D$ & $D^*$ & $D_s$ & $D_s^*$ &$\qquad$ & $\eta_c$ &$J/\psi$
  \\
\cline{1-5}\cline{7-8}
  $f_{P/V}$ &  215(11) & 224(18)  &255.7(8.9) &277(11) &   &383.8(6.7) & -\\
 $f_V^T$ &  - & 191(14) &-   &252(10) &  &-  & 391.1(5.4)  \\
\cline{1-5}\cline{7-8}

\end{tabular}
\end{table}

\begin{table}[!htbp]
    \caption{Comparisons of $f_{D_{(s)}^*}$ (in MeV). }
    \label{comparison}
    \centering
   \setlength{\tabcolsep}{4pt}
  \renewcommand{\arraystretch}{1.3}
\begin{tabular}{ccccccc}
\hline 
   &   [This work] & \cite{Chen:2020qma} & \cite{Donald:2013sra}  & \cite{Blossier:2018jol} & \cite{Lubicz:2017asp} & \cite{Becirevic:2012ti}
  \\
\hline
  $f_{D^*}$ &  224(18)  & 234(6)   & -  &  -  & 223.5(8.4) & 278(17)   \\
 $f_{D_s^*}$ &  277(11)  &  274(7)   &  274(6) &  264(15) & 268.8(6.6) & 311(9)   \\
\hline

\end{tabular}
\end{table}

By using our result of $f_{D_s^*}=277(11)$ MeV, one can obtain the decay width of the pure leptonic decay of $D_s^*$, which is $\Gamma(D_s^*\to \ell\nu_\ell)|_{\ell=e,\mu}=2.5(2)\times10^{-6}$ keV, as HPQCD did in~\cite{Donald:2013sra}. Combining this result with the total decay width of $D_s^*$, which is $\Gamma_{\rm tot}(D_s^*)=$0.0587(54) keV~\cite{Meng:2024gpd}, we then find the branching ratio
\begin{equation}
    {\rm Br}(D_s^*\to \ell\nu_\ell)|_{\ell=e,\mu}=4.26(52)\times 10^{-5}.
\end{equation}
This value can be confronted with experiments in the future.


Discretization effects in the decay constants of charmed mesons are found to be larger than those in the meson masses. As for the decay constants of charmonia, the discretization effects are even larger. While for ratios of decay constants, the lattice artefacts are much smaller. 

Currently, our estimation of discretization errors are based on the analyses of results from two sets of lattice with inverse lattice spacing of $1.730\,$GeV and $2.383\,$GeV. To remove the discretization errors, we need to repeat our calculations on other sets of lattice with different spacings. Additionally, QED corrections and contributions from the QCD-disconnected contractions for charmonia are necessary for high-precision researches in the future.

\section*{ACKNOWLEDGMENTS}
We thank the RBC-UKQCD collaborations for sharing the domain wall fermion configurations.
This work is supported in part by National Key Research and Development Program of China under Contract No. 2020YFA0406400, No. 2023YFA1606002 and by the National Natural Science Foundation of China (NSFC) under Grants No. 12075253, No. 11935017, No. 12192264, No. 12293060, No. 12293065, No. 12293063 and 12070131001 (CRC 110 by DFG and NNSFC).
K. L. is supported by the U.S. DOE Grant No. DE-SC0013065 and DOE Grant No. DEAC05-06OR23177, which is within the framework of the TMD Topical Collaboration.
The GWU code~\cite{Alexandru:2011ee,Alexandru:2011sc} is acknowledged.
The computations were performed on the HPC clusters at Institute of High Energy Physics (Beijing).


\begin{thebibliography}{99}


\bibitem{ParticleDataGroup:2022pth}
R.~L.~Workman \textit{et al.} [Particle Data Group],
PTEP \textbf{2022}, 083C01 (2022)
doi:10.1093/ptep/ptac097

\bibitem{Dobrescu:2008er}
B.~A.~Dobrescu and A.~S.~Kronfeld,
Phys. Rev. Lett. \textbf{100}, 241802 (2008)
doi:10.1103/PhysRevLett.100.241802
[arXiv:0803.0512 [hep-ph]].

\bibitem{FlavourLatticeAveragingGroupFLAG:2021npn}
Y.~Aoki \textit{et al.} [Flavour Lattice Averaging Group (FLAG)],
Eur. Phys. J. C \textbf{82}, no.10, 869 (2022)
doi:10.1140/epjc/s10052-022-10536-1
[arXiv:2111.09849 [hep-lat]].

\bibitem{FermilabLattice:2011njy}
A.~Bazavov \textit{et al.} [Fermilab Lattice and MILC],
Phys. Rev. D \textbf{85} (2012), 114506
doi:10.1103/PhysRevD.85.114506
[arXiv:1112.3051 [hep-lat]].

\bibitem{Na:2012iu}
H.~Na, C.~T.~H.~Davies, E.~Follana, G.~P.~Lepage and J.~Shigemitsu,
Phys. Rev. D \textbf{86} (2012), 054510
doi:10.1103/PhysRevD.86.054510
[arXiv:1206.4936 [hep-lat]].

\bibitem{Carrasco:2014poa}
N.~Carrasco, P.~Dimopoulos, R.~Frezzotti, P.~Lami, V.~Lubicz, F.~Nazzaro, E.~Picca, L.~Riggio, G.~C.~Rossi and F.~Sanfilippo, \textit{et al.}
Phys. Rev. D \textbf{91} (2015) no.5, 054507
doi:10.1103/PhysRevD.91.054507
[arXiv:1411.7908 [hep-lat]].

\bibitem{Boyle:2017jwu}
P.~A.~Boyle, L.~Del Debbio, A.~J\"uttner, A.~Khamseh, F.~Sanfilippo and J.~T.~Tsang,
JHEP \textbf{12} (2017), 008
doi:10.1007/JHEP12(2017)008
[arXiv:1701.02644 [hep-lat]].

\bibitem{Bazavov:2017lyh}
A.~Bazavov, C.~Bernard, N.~Brown, C.~Detar, A.~X.~El-Khadra, E.~G\'amiz, S.~Gottlieb, U.~M.~Heller, J.~Komijani and A.~S.~Kronfeld, \textit{et al.}
Phys. Rev. D \textbf{98} (2018) no.7, 074512
doi:10.1103/PhysRevD.98.074512
[arXiv:1712.09262 [hep-lat]].

\bibitem{Boyle:2018knm}
P.~A.~Boyle \textit{et al.} [RBC/UKQCD],
[arXiv:1812.08791 [hep-lat]].


\bibitem{Bussone:2023kag}
A.~Bussone \textit{et al.} [Alpha],
Eur. Phys. J. C \textbf{84}, no.5, 506 (2024)
doi:10.1140/epjc/s10052-024-12816-4
[arXiv:2309.14154 [hep-lat]].

\bibitem{Kuberski:2024pms}
S.~Kuberski, F.~Joswig, S.~Collins, J.~Heitger and W.~S\"oldner,
[arXiv:2405.04506 [hep-lat]].

\bibitem{BESIII:2023zjq}
M.~Ablikim \textit{et al.} [BESIII],
Phys. Rev. Lett. \textbf{131}, no.14, 141802 (2023)
doi:10.1103/PhysRevLett.131.141802
[arXiv:2304.12159 [hep-ex]].

\bibitem{Cheng:2022mvd}
S.~Cheng, Y.~h.~Ju, Q.~Qin and F.~s.~Yu,
Eur. Phys. J. C \textbf{82}, no.11, 1037 (2022)
doi:10.1140/epjc/s10052-022-10987-6
[arXiv:2203.06797 [hep-ph]].

\bibitem{Yang:2022ece}
Y.~Yang, K.~Li, Z.~Li, J.~Huang, Q.~Chang and J.~Sun,
Phys. Rev. D \textbf{106}, no.3, 036029 (2022)
doi:10.1103/PhysRevD.106.036029
[arXiv:2207.10277 [hep-ph]].

\bibitem{Cui:2023jiw}
B.~Y.~Cui, Y.~K.~Huang, Y.~M.~Wang and X.~C.~Zhao,
Phys. Rev. D \textbf{108}, no.7, L071504 (2023)
doi:10.1103/PhysRevD.108.L071504
[arXiv:2301.12391 [hep-ph]].

\bibitem{Zhou:2015jba}
S.~H.~Zhou, Y.~B.~Wei, Q.~Qin, Y.~Li, F.~S.~Yu and C.~D.~Lu,
Phys. Rev. D \textbf{92}, no.9, 094016 (2015)
doi:10.1103/PhysRevD.92.094016
[arXiv:1509.04060 [hep-ph]].

\bibitem{Donald:2013sra}
G.~C.~Donald, C.~T.~H.~Davies, J.~Koponen and G.~P.~Lepage,
Phys. Rev. Lett. \textbf{112}, 212002 (2014)
doi:10.1103/PhysRevLett.112.212002
[arXiv:1312.5264 [hep-lat]].

\bibitem{Becirevic:2012ti}
D.~Becirevic, V.~Lubicz, F.~Sanfilippo, S.~Simula and C.~Tarantino,
JHEP \textbf{02}, 042 (2012)
doi:10.1007/JHEP02(2012)042
[arXiv:1201.4039 [hep-lat]].

\bibitem{Blossier:2018jol}
B.~Blossier, J.~Heitger and M.~Post,
Phys. Rev. D \textbf{98}, no.5, 054506 (2018)
doi:10.1103/PhysRevD.98.054506
[arXiv:1803.03065 [hep-lat]].

\bibitem{Lubicz:2017asp}
V.~Lubicz \textit{et al.} [ETM],
Phys. Rev. D \textbf{96}, no.3, 034524 (2017)
doi:10.1103/PhysRevD.96.034524
[arXiv:1707.04529 [hep-lat]].

\bibitem{Gambino:2019vuo}
P.~Gambino, V.~Lubicz, A.~Melis and S.~Simula,
J. Phys. Conf. Ser. \textbf{1137}, no.1, 012005 (2019)
doi:10.1088/1742-6596/1137/1/012005

\bibitem{Gelhausen:2013wia}
P.~Gelhausen, A.~Khodjamirian, A.~A.~Pivovarov and D.~Rosenthal,
Phys. Rev. D \textbf{88}, 014015 (2013)
[erratum: Phys. Rev. D \textbf{89}, 099901 (2014); erratum: Phys. Rev. D \textbf{91}, 099901 (2015)]
doi:10.1103/PhysRevD.88.014015
[arXiv:1305.5432 [hep-ph]].

\bibitem{Wang:2015mxa}
Z.~G.~Wang,
Eur. Phys. J. C \textbf{75}, 427 (2015)
doi:10.1140/epjc/s10052-015-3653-9
[arXiv:1506.01993 [hep-ph]].

\bibitem{BESIII:2022wsp}
M.~Ablikim \textit{et al.} [BESIII],
[arXiv:2206.13674 [hep-ex]].

\bibitem{Babiarz:2019sfa}
I.~Babiarz, V.~P.~Goncalves, R.~Pasechnik, W.~Sch\"afer and A.~Szczurek,
Phys. Rev. D \textbf{100}, no.5, 054018 (2019)
doi:10.1103/PhysRevD.100.054018
[arXiv:1908.07802 [hep-ph]].

\bibitem{Ryu:2018egt}
H.~Y.~Ryu, H.~M.~Choi and C.~R.~Ji,
Phys. Rev. D \textbf{98}, no.3, 034018 (2018)
doi:10.1103/PhysRevD.98.034018
[arXiv:1804.08287 [hep-ph]].

\bibitem{Geng:2013yfa}
C.~Q.~Geng and C.~C.~Lih,
Eur. Phys. J. C \textbf{73}, no.8, 2505 (2013)
doi:10.1140/epjc/s10052-013-2505-8
[arXiv:1307.3852 [hep-ph]].

\bibitem{Dudek:2006ej}
J.~J.~Dudek, R.~G.~Edwards and D.~G.~Richards,
Phys. Rev. D \textbf{73}, 074507 (2006)
doi:10.1103/PhysRevD.73.074507
[arXiv:hep-ph/0601137 [hep-ph]].

\bibitem{Becirevic:2013bsa}
D.~Be\v{c}irevi\'c, G.~Duplan\v{c}i\'c, B.~Klajn, B.~Meli\'c and F.~Sanfilippo,
Nucl. Phys. B \textbf{883}, 306-327 (2014)
doi:10.1016/j.nuclphysb.2014.03.024
[arXiv:1312.2858 [hep-ph]].

\bibitem{Bailas:2018car}
G.~Bailas, B.~Blossier and V.~Mor\'enas,
Eur. Phys. J. C \textbf{78}, no.12, 1018 (2018)
doi:10.1140/epjc/s10052-018-6495-4
[arXiv:1803.09673 [hep-lat]].

\bibitem{Davies:2010ip}
C.~T.~H.~Davies, C.~McNeile, E.~Follana, G.~P.~Lepage, H.~Na and J.~Shigemitsu,
Phys. Rev. D \textbf{82}, 114504 (2010)
doi:10.1103/PhysRevD.82.114504
[arXiv:1008.4018 [hep-lat]].

\bibitem{Donald:2012ga}
G.~C.~Donald, C.~T.~H.~Davies, R.~J.~Dowdall, E.~Follana, K.~Hornbostel, J.~Koponen, G.~P.~Lepage and C.~McNeile,
Phys. Rev. D \textbf{86}, 094501 (2012)
doi:10.1103/PhysRevD.86.094501
[arXiv:1208.2855 [hep-lat]].

\bibitem{Hatton:2020qhk}
D.~Hatton \textit{et al.} [HPQCD],
Phys. Rev. D \textbf{102}, no.5, 054511 (2020)
doi:10.1103/PhysRevD.102.054511
[arXiv:2005.01845 [hep-lat]].

\bibitem{Chen:2020qma}
Y.~Chen \textit{et al.} [\ensuremath{\chi}QCD],
Chin. Phys. C \textbf{45}, no.2, 023109 (2021)
doi:10.1088/1674-1137/abcd8f
[arXiv:2008.05208 [hep-lat]].

\bibitem{RBC:2010qam}
Y.~Aoki \textit{et al.} [RBC and UKQCD],
Phys. Rev. D \textbf{83}, 074508 (2011)
doi:10.1103/PhysRevD.83.074508
[arXiv:1011.0892 [hep-lat]].

\bibitem{RBC:2014ntl}
T.~Blum \textit{et al.} [RBC and UKQCD],
Phys. Rev. D \textbf{93}, no.7, 074505 (2016)
doi:10.1103/PhysRevD.93.074505
[arXiv:1411.7017 [hep-lat]].

\bibitem{Neuberger:1997fp}
H.~Neuberger,
Phys. Lett. B \textbf{417}, 141-144 (1998)
doi:10.1016/S0370-2693(97)01368-3
[arXiv:hep-lat/9707022 [hep-lat]].

\bibitem{Chiu:1998gp}
T.~W.~Chiu and S.~V.~Zenkin,
Phys. Rev. D \textbf{59}, 074501 (1999)
doi:10.1103/PhysRevD.59.074501
[arXiv:hep-lat/9806019 [hep-lat]].

\bibitem{DeGrand:2005af}
T.~A.~DeGrand and Z.~f.~Liu,
Phys. Rev. D \textbf{72}, 054508 (2005)
doi:10.1103/PhysRevD.72.054508
[arXiv:hep-lat/0507017 [hep-lat]].

\bibitem{Liu:2013yxz}
Z.~Liu \textit{et al.} [chiQCD],
Phys. Rev. D \textbf{90}, no.3, 034505 (2014)
doi:10.1103/PhysRevD.90.034505
[arXiv:1312.7628 [hep-lat]].
\bibitem{Bi:2017ybi}
Y.~Bi, H.~Cai, Y.~Chen, M.~Gong, K.~F.~Liu, Z.~Liu and Y.~B.~Yang,
Phys. Rev. D \textbf{97}, no.9, 094501 (2018)
doi:10.1103/PhysRevD.97.094501
[arXiv:1710.08678 [hep-lat]].
\bibitem{He:2022lse}
F.~He \textit{et al.} [\ensuremath{\chi}QCD],
Phys. Rev. D \textbf{106}, no.11, 114506 (2022)
doi:10.1103/PhysRevD.106.114506
[arXiv:2204.09246 [hep-lat]].

\bibitem{Bi:2023pnf}
Y.~Bi \textit{et al.} [\ensuremath{\chi}QCD],
Phys. Rev. D \textbf{108}, no.5, 5 (2023)
doi:10.1103/PhysRevD.108.054506
[arXiv:2302.01659 [hep-lat]].

\bibitem{FermilabLatticeHPQCD:2023jof}
A.~Bazavov \textit{et al.} [Fermilab Lattice, HPQCD, and MILC],
Phys. Rev. D \textbf{107}, no.11, 114514 (2023)
doi:10.1103/PhysRevD.107.114514
[arXiv:2301.08274 [hep-lat]].

\bibitem{Dowdall:2013rya}
R.~J.~Dowdall, C.~T.~H.~Davies, G.~P.~Lepage and C.~McNeile,
Phys. Rev. D \textbf{88}, 074504 (2013)
doi:10.1103/PhysRevD.88.074504
[arXiv:1303.1670 [hep-lat]].

\bibitem{Lepage:2001ym}
G.~P.~Lepage \textit{et al.} [HPQCD],
Nucl. Phys. B Proc. Suppl. \textbf{106}, 12-20 (2002)
doi:10.1016/S0920-5632(01)01638-3
[arXiv:hep-lat/0110175 [hep-lat]].

\bibitem{Du:2024wtr}
H.~Y.~Du, B.~Hu, Y.~Chen, H.~T.~Ding, C.~Liu, L.~Liu, Y.~Meng, P.~Sun, J.~H.~Wang and Y.~B.~Yang, \textit{et al.}
[arXiv:2408.03548 [hep-lat]].

\bibitem{Borsanyi:2020mff}
S.~Borsanyi, Z.~Fodor, J.~N.~Guenther, C.~Hoelbling, S.~D.~Katz, L.~Lellouch, T.~Lippert, K.~Miura, L.~Parato and K.~K.~Szabo, \textit{et al.}
Nature \textbf{593}, no.7857, 51-55 (2021)
doi:10.1038/s41586-021-03418-1
[arXiv:2002.12347 [hep-lat]].

\bibitem{Aoki:2007xm}
Y.~Aoki, P.~A.~Boyle, N.~H.~Christ, C.~Dawson, M.~A.~Donnellan, T.~Izubuchi, A.~Juttner, S.~Li, R.~D.~Mawhinney and J.~Noaki, \textit{et al.}
Phys. Rev. D \textbf{78}, 054510 (2008)
doi:10.1103/PhysRevD.78.054510
[arXiv:0712.1061 [hep-lat]].

\bibitem{Davies:2009tsa}
C.~T.~H.~Davies \textit{et al.} [HPQCD],
Phys. Rev. D \textbf{81}, 034506 (2010)
doi:10.1103/PhysRevD.81.034506
[arXiv:0910.1229 [hep-lat]].

\bibitem{Hatton:2020vzp}
D.~Hatton \textit{et al.} [HPQCD],
Phys. Rev. D \textbf{102}, no.9, 094509 (2020)
doi:10.1103/PhysRevD.102.094509
[arXiv:2008.02024 [hep-lat]].

\bibitem{Meng:2024gpd}
Y.~Meng, J.~L.~Dang, C.~Liu, Z.~Liu, T.~Shen, H.~Yan and K.~L.~Zhang,
Phys. Rev. D \textbf{109}, no.7, 074511 (2024)
doi:10.1103/PhysRevD.109.074511
[arXiv:2401.13475 [hep-lat]].

\bibitem{Alexandru:2011ee}
A.~Alexandru, C.~Pelissier, B.~Gamari and F.~Lee,
J. Comput. Phys. \textbf{231}, 1866-1878 (2012)
doi:10.1016/j.jcp.2011.11.003
[arXiv:1103.5103 [hep-lat]].

\bibitem{Alexandru:2011sc}
A.~Alexandru, M.~Lujan, C.~Pelissier, B.~Gamari and F.~X.~Lee,
doi:10.1109/SAAHPC.2011.13
[arXiv:1106.4964 [hep-lat]].



\end{thebibliography}
\end{document}